\RequirePackage{amsmath}

\documentclass[12pt]{iopart}
\pdfoutput=1
\usepackage{graphicx}
\usepackage{amsmath,amsfonts,amssymb}

\usepackage{xcolor}

\colorlet{mylinkcolor}{blue!66!black!80}

\usepackage[colorlinks=true,linkcolor=mylinkcolor,citecolor=mylinkcolor,filecolor=cyan,urlcolor=mylinkcolor,breaklinks=true]{hyperref}

\usepackage[utf8]{inputenc}

\usepackage{cite}

\newcommand{\matr}[1]{\textbf{#1}}

\newcommand{\del}{\partial}
\newcommand{\dd}{\mathrm{d}}

\newcommand{\mL}{\mathbf{L}}
\newcommand{\mLa}{\mathbf{L}_a}

\newcommand{\T}{\top}

\newcommand{\bra}[1]{\langle#1|}
\newcommand{\ket}[1]{|#1\rangle}
\newcommand{\kket}[1]{#1\rangle}
\newcommand{\id}{\mathbf{1 }}
\newcommand{\adj}{\operatorname{adj}}

\newcommand{\psiL}{\psi^\mathrm{L}}
\newcommand{\psiR}{\psi^\mathrm{R}}
\newcommand{\phiL}{\phi^\mathrm{L}}
\newcommand{\phiR}{\phi^\mathrm{R}}

\newcommand{\Nmax}{M}

\newcommand{\kb}{k_\mathrm{B}}

\newcommand{\mysubref}[2]{\href{#1}{\ref{#1}#2}}

\newcommand{\mcA}{\boldsymbol{\mathcal{A}}}

\newcommand{\cA}{\boldsymbol{\mathcal{A}}}

\begin{document}
 \title[Interlacing Relaxation and First-Passage Phenomena]{Interlacing Relaxation and First-Passage Phenomena in Reversible Discrete and Continuous Space Markovian Dynamics
}
 \author{David Hartich and Alja\v{z} Godec}
\address{Mathematical Biophysics Group, Max-Planck-Institute for
  Biophysical Chemistry, G\"{o}ttingen 37077, Germany}
\eads{\href{mailto:david.hartich@mpibpc.mpg.de}{david.hartich@mpibpc.mpg.de},\href{mailto:agodec@mpibpc.mpg.de}{agodec@mpibpc.mpg.de}}

\begin{abstract}
We uncover a duality between relaxation and first passage processes in ergodic reversible Markovian dynamics in both discrete and continuous state-space. The duality exists in the form of a spectral interlacing -- the respective time scales of relaxation and first passage are shown to interlace. Our canonical theory allows for the first time to determine the full first passage time distribution analytically from the simpler relaxation eigenspectrum. The duality is derived and proven rigorously for both discrete state Markov processes in arbitrary dimension and effectively one-dimensional diffusion processes, whereas we also discuss extensions to more complex scenarios. We apply our theory to a simple discrete-state protein folding model and to the Ornstein-Uhlenbeck process, for which we obtain the exact first passage time distribution analytically in terms of a Newton series of determinants of almost triangular matrices.
\end{abstract}


\date{\today}

\section{Introduction}

In his seminal work  \cite{kram40} Kramers analyzed the kinetics of
chemical reactions in terms of diffusive barrier crossing, assuming that
the kinetic rate of a chemical reaction
corresponds to the inverse of the mean first crossing time. 
Ever since, first passage theory is at the heart of theoretical
descriptions of kinetics of chemical reactions 
\cite{szab80,ben93,osha95,meji11,guer16,li17}; see e.g. \cite{haen90,redn01,metz14a,beni14} for comprehensive reviews.
 
In a broader context, first passage concepts were invoked
in studies of kinetics in complex media, such as
reactions in fractal-like \cite{kope88,avra00} and planar domains \cite{rupp15,greb16},
in inhomogeneous cellular environments
\cite{bres13,gode15,vacc15,gode16a}, in the study of
neural networks \cite{vile09,brau15}, ultra cold atoms \cite{bark14},
 as well as in diverse narrow escape problems
\cite{sing06,schu07,rein09,pill10,isaa16,greb17} and so-called intermittent search strategies involving searching agents with internal dynamics
\cite{paly14,gode17} (see also \cite{beni11} for a review).

First passage times play an important role in quantifying persistence properties
in non-equilibrium interacting many-body systems
\cite{maju01,maju02,bray13}. More recent applications of first passage
concepts also include stochastic thermodynamics \cite{jarz11,seif12,broe15},
in particular, fluctuation relations for stopping time statistics and stochastic entropy production in
driven molecular systems
\cite{neri17} and in stochastic resetting processes \cite{fuch16,rold17},
as well as uncertainty relations for first passage time statistics of fluctuating currents
\cite{garr17,ging17} (see also \cite{bara15}).

Moreover, our current understanding of the speed and precision of
transcription regulation in biological cells, and in particular of the
role of the so-called proximity effect in the co-regulation of genes,
\cite{kole07,fras07} builds on first passage time ideas. 
The corresponding physical principles underlying these proximity effects
were explained in \cite{beni10,meye11,gode16}. Notably, these works
revealed the inherent insufficiency of the mean first passage time
and traditional rate-based concepts  for a quantitative description of biophysical dynamics
in the so-called few encounter limit \cite{gode16}. As a result,
a quantitative understanding of phenomena such as gene regulation \cite{kole07,fras07,beni10,meye11,bial12,pulk13,gode16} and the misfolding-triggered
pathological aggregation of proteins  \cite{dobs03,chit06,zhen13, yu15,dee16,nguy17}, which
are discussed in more detail in a related study \cite{hart18_arxiv}, requires the consideration of
the full statistics of first passage time.

Existing studies of the full first passage statistics in physical systems
typically focus on systems with continuous state-space dynamics,
whereas much less emphasis is put on discrete-space dynamics
\cite{schn17a}. Recent investigations of such discrete-state dynamics
include, for example, simple models of enzyme kinetics \cite{muns09,bel10,grim17}
and novel numerical approximation schemes for studying  first-passage statistics based on Bayesian inference
 \cite{schn17} (see also \cite{webe17} for a recent review).

Complementary to first passage processes are relaxation dynamics,
which by contrast do not terminate upon reaching a given threshold
for the first time. Relaxation phenomena in reversible diffusive
dynamical systems are nowadays well understood in terms of the eigenmodes and eigenvalues
of the underlying Fokker-Planck operators, which provide a generic and
very intuitive understanding of the dynamics of complex stochastic systems
\cite{biro01,tana03,tana04}. Conversely, despite for allowing an
analogous spectral representation, a similar intuitive understanding of the
full first passage statistics and its physical implications remains
elusive. Notwithstanding, an important approximate link between 
the mean first passage time for escaping the deepest potential
basin and the corresponding slowest relaxation mode in the potential
was established in the seminal works of Matkovsky and Schuss
\cite{schu79,matk81}, which has ever since been used routinely in
explaining relaxation phenomena in condensed matter
systems. Nevertheless, a deeper and more generic connection between
the two paradigms to date was not yet established. 

Here, we present the complete duality between relaxation and first
passage phenomena, which holds for all ergodic Markov
processes obeying detailed balance in both, continuous and discrete 
state-space, in which the absorbing target is effectively one-dimensional. The duality emerges in the form of a spectral
interlacing, which we prove rigorously by combining spectral-theoretic, matrix-algebraic
and Greens function-theoretic concepts.
On the one hand the duality
allows for an intuitive generic understanding of first passage
phenomena in terms of relaxation eigenmodes. On the other hand, it
enables us to determine the full first passage time statistics exactly
from the corresponding relaxation eigensystem.  The formalism is exact
and holds for all reversible Markovian systems governed by a master
equation in arbitrary dimensions or by a Fokker-Planck equation, and
therefore unifies the theoretical treatment of discrete and
continuous space phenomena. 
We note the spectral interlacing in the case of a discrete state dynamics
has also recently be deduced from a `lumping' of the state dynamics \cite{gron08}.

To illustrate the predictive power of the formalism in practice, we
here predominantly focus on systems with discrete state-space dynamics,
whereas continuous space dynamics are treated in more detail in a related study \cite{hart18_arxiv}. In particular, we here apply our theory to a simple discrete-state protein folding model
and to diffusion in a harmonic potential, also know as the
Ornstein-Uhlenbeck process. Notably, we obtain, to the best of our
knowledge, for the first time an exact analytical solution for the full first passage time distribution of the Ornstein-Uhlenbeck process in the time domain.

The paper is organized as follows. In Sec.~\ref{sec:def_ME_vs_FPE}
we present a canonical formulation the first passage problem
applicable to both discrete states-pace and continuous Fokker-Planck
dynamics. Sections \ref{sec:main_discrete} and \ref{sec:main_cont}
provide a step-by-step explanation of how one can exactly determine
the first-passage distribution from the corresponding relaxation process, and also contain rigorous proofs of the duality in
discrete and continuous state-space dynamics, respectively.
We apply the duality framework in Sec.~\ref{sec:examples} to determine the
first passage statistics for a simple
protein folding model and for the Ornstein-Uhlenbeck process. A concluding
perspective is provided in Sec.~\ref{sec:conclusion}.
In \ref{sec:A:mu1} we derive a compact representation of the long-time
asymptotics of the first passage time distribution, which \emph{inter alia}
extends our results for the long time asymptotics from equilibrium systems to irreversibly driven systems.

\section{Fundamentals}
\label{sec:def_ME_vs_FPE}
\subsection{Relaxation and first passage}
We assume that the probability density to find the system in state
$x$ at time $t$ upon evolving from an initial state $x_0$ according to
microscopically reversible dynamics, $P(x,t|x_0)$, is governed
by
\begin{equation}
 \del_t P(x,t|x_0)=\mL P(x,t|x_0),
 \label{eq:1}
\end{equation}
where $\mL$ is a linear reversible operator, which will be specified
below. We consider two classes of operators: (DS)~discrete state Markov jump
process, where $x$ and $x_0$ assume only a finite number of states,
and (FP)~continuous Markovian diffusion governed by a Fokker-Planck equation.

For discrete Markov state models of class (DS)
the dynamics is governed by
\begin{equation}
  \mL P(x,t|x_0)\equiv\sum_{x'=0}^\Nmax L(x,x') P(x',t|x_0),
  \label{eq:MEgen}
\end{equation}
where $x,x'=0,1,\ldots,\Nmax$ denote the discrete states, $L(x,x')$ is
the rate of jumping from state $x'$ to state $x$  ($x\neq x'$) and
$-L(x,x)=\sum_{x'\neq x} L(x',x)$ is the total rate of leaving state $x$ guaranteeing conservation
of probability ($\sum_x\del_t P(x,t|x_0)=0$).
In order to have reversible dynamics
we need to
additionally impose detailed balance, i.e. the constraint
$L(x,x')/L(x',x)=\exp[\beta U(x')-\beta U(x)]$ (see., e.g.,
\cite{kamp07}), which assures that the system will relax to a
Boltzmann distribution in a potential $U(x)$ on ergodic timescales
$P_\mathrm{eq}(x)\propto \e^{-\beta U(x)}$, where $\beta=1/k_\mathrm{B} T$ is the
inverse thermal energy.
We call such a reversible ergodic process that conserves probability a \emph{relaxation process}.
If we add an absorbing point at $x=a$ we call the resulting process
a \emph{first passage process} or in short absorption, which we introduce in the following way.
First, we modify the generator ($\mL\to \mLa$) such that all transitions corresponding
to jumps out of the absorbing state $a$ are removed, i.e., the
elements of the
first passage generator read
\begin{equation}
 L_a(x,x')=\left\{
 \begin{array}{ll}
    0&\text{if $x'=a$,}\\
  L(x,x')&\text{otherwise}.
 \end{array}
\right.
 \label{eq:La_elementwise}
\end{equation}
Using a bra-ket matrix notation \cite{kada68} we rewrite this equation as
\begin{equation}
 \mLa=\mL-\mL\ket{a}\bra{a},
 \label{eq:La_def}
\end{equation}
where $\ket{a}=(0,\ldots ,0,1,0,\ldots,0)^\T=(\ket{a})^\T$ is a vector with all entries except the $a$th one; consequently,
we identify $L_a(x,x')=\bra{x}\mLa\ket{x'}$.
The first passage time density to reach state $a$ at time $t$
starting from $x_0$
is then formally defined by
\begin{equation}
 \wp_a(t|x_0)=\del_t\bra{a}\e^{\mLa t}\ket{x_0}=\bra{a}\mLa \e^{\mLa t}\ket{x_0},
 \label{eq:pFP_def}
\end{equation}
which is nothing but the normalized probability flux into state $a$
with
$\int_0^\infty\wp_a(t|x_0)\dd t=1$.
Note that with Eq.~\eqref{eq:La_def} we use the convention that
$\ket{a}$ is the unique stationary solution with $\mLa\ket{a}=0$.

For a continuous space Markovian diffusion the transition probability density
function (the `propagator') instead obeys the Fokker-Planck equation \eqref{eq:1} 
\begin{align}
 \mL P(x,t|x_0)&=-\del_x j(x,t|x_0)\nonumber\\
 &\equiv\del_x D[\beta U'(x)+\del_x]P(x,t|x_0),
 \label{eq:FPgen}
\end{align}
where $j(x,t|x_0)$ is the probability current, $D$ is the diffusion
constant, $-U'(x)=-\del_xU(x)$ is a force field generated by the potential $U(x)$
at position $x$, and $\beta$ is the inverse temperature, which we set to $\beta\equiv 1$ to express energies in units of $\kb T$ from now on.
The scenario with reflecting barriers at $x=b_\pm$ with $j(b_\pm,t|x_0)=0$
\cite{redn01}, we term a \emph{relaxation process}, where
$b_\pm=\pm\infty$ correspond to so-called natural boundary conditions~\footnote{For natural boundary conditions the current and the probability density both vanish, i.e., $\lim_{b\to\pm\infty}j(b,t|x_0)=\lim_{b\to\pm\infty}P(b,t|x_0)=0$.
}.

Conversely, an absorbing boundary at $x=a$ enters the Fokker-Planck equation via the Dirichlet boundary condition $P(a,t|x_0)=0$,
without altering the partial differential equation \eqref{eq:FPgen}, i.e.,
the first passage operator still reads $\mLa=\del_xD[U'(x)+\del_x]$.
However, here the first passage time density becomes the probability flux into state $a$.
For convenience we use the operator $\mLa$ as shorthand for
Eq.~\eqref{eq:FPgen} under the boundary condition
$P(a,t|x_0)=0$.

We note that without an absorbing point both, dynamics governed by the master equation \eqref{eq:MEgen} and the Fokker-Planck equation \eqref{eq:FPgen}
relax to the Boltzmann distribution $P_{\mathrm{eq}}(x)\propto \exp[-U(x)]$, whereas
with the absorbing boundary condition the particle will eventually reach the target
with probability 1.

\subsection{Eigendecomposition}
\label{sec:Eigen_decomposition}
Since $\mL$ is assumed to generate a reversible Markov process,
we can expand the generator
$\mL$ in a bi-orthogonal eigenbasis \cite{gard04}. Denoting the
eigenvalues of the relaxation process by $\lambda_k$ and the
corresponding left (right) eigenvectors by $\bra{\psiL_k}$
($\ket{\psiR_k}$), respectively, the generators from
Eqs.~\eqref{eq:MEgen} and \eqref{eq:FPgen} 
become in the respective eigenbases
\begin{equation}
 \mL =-\sum_k \lambda_k \ket{\psiR_k}\bra{\psiL_k},
 \label{eq:decomp_relax}
\end{equation}
where $\lambda_0=0\le\lambda_1\le\ldots$, and $\bra{\psiL_k}\kket{\psiR_l}=\delta_{kl}$.
We assume the eigenvalues to be ordered
such that $\lambda_k\le\lambda_{k+1}$, and the generator to be irreducible
$\lambda_0=0<\lambda_1$, which means that there is a unique equilibrium state \cite{kamp07}.
Note that for a Fokker-Planck
equation with reflecting barriers at $x=b_\pm$ (relaxation) the eigenfunction $\psiR_k(x)\equiv\bra{x}\kket{\psiR_k}$ must satisfy the zero flux condition $-D[U'(x)+\del_x]\psiR_k(x)|_{x=b_\pm}=0$, with $\bra{x}\kket{\psiR_0}\propto\e^{-U(x)}$.

The generator with the absorbing point at state $a$, can similarly be expanded in a bi-orthogonal set of eigenfunctions
\begin{equation}
 \mLa=-\sum_k \mu_k \ket{\phiR_k}\bra{\phiL_k},
  \label{eq:decomp_FP}
\end{equation}
where $\mu_k$ is the $k$-th eigenvalue and $\bra{\phiL_k}$ ($\ket{\phiR_k}$) denote the corresponding left (right) eigenfunctions
of the first passage process.
Without loss generality we use an ordered labeling such that $\mu_k\le\mu_{k+1}$,
where $0<\mu_1$.

The left and right eigenvectors of the absorption (at position $x\neq
a$) as well as of the relaxation process
are related via $\bra{x}\kket{\phiL_k}\propto\e^{U(x)}\bra{x}\kket{\phiR_k}$ and
$\bra{x}\kket{\psiL_k}\propto\e^{U(x)}\bra{x}\kket{\psiR_k}$, respectively.
In the case of a discrete number of states, the lowest eigenvalue of the generator \eqref{eq:La_def}
will be $\mu_0=0$ with the right eigenfunction $\ket{\phiR_0}=\ket{a}$,
whereas for Fokker-Planck dynamics one imposes the boundary condition
$\bra{a}\kket{\phiR_k}=0$.

In a previous work an explicit Newton series expression for $\mu_1$ in
terms of a series of almost triangular matrices was derived \cite{gode16}, which corresponds to a large deviation limit $t\to\infty$.
One of our main goals here is to obtain the full first passage statistics $\wp_a(t|x_0)$ explicitly
in terms of relaxation eigenmodes. Our theory builds on the renewal
theorem, which we briefly review
in the following subsection.

\subsection{Renewal theorem}
The classical renewal theorem provides
a well known implicit connection between first passage
and relaxation processes. It relates the 
probability density of the freely propagating system to be in state $x$ at time $t$
upon starting from a state $x_0$, to the first passage distribution
$\wp_a(t|x_0)$ from $x_0$ to $a$:
\begin{equation}
  P(x,t|x_0)=\int_0^t\dd \tau  P(x,t-\tau|a) \wp_a(\tau|x_0),
  \label{eq:renewal}
\end{equation}
where both $ P(x,t|x_0)$ and $\wp_a(t|x_0)$ admit a spectral representation
\begin{equation}
 P(x,t|x_0)=\bra{x}\e^{\mL t}\ket{x_0}=\sum_k
 \bra{x}\kket{\psiR_k}\bra{\psiL_k}\kket{x_0}\e^{-\lambda_kt}\label{eq:propagator}\end{equation}
and
\begin{equation}
 \wp_a(t|x_0)=\sum_{k\ge 1} w_k(x_0)\mu_k\e^{-\mu_kt }, \label{eq:FPTdensity_def}
\end{equation}
respectively. In other words, a system starting from state $x_0$ must pass through state $a$
before reaching  the final state $x$,
which for an effectively 1-dimensional Fokker-Planck necessarily means $x_0< a\le x$ or $x_0> a\ge x$. In
Eq.~\eqref{eq:FPTdensity_def} we introduced in the first passage weights
\begin{equation}
w_k(x_0)=
\left\{\begin{array}{ll}
 -\bra{a}\kket{\phiR_k} \bra{\phiL_k}\kket{x_0}&\text{for DS},\\ 
-\sigma_\pm D\displaystyle{\frac{\del\bra{x}\kket{\phiR_k}}{\del x}\Big|_{x=a} \frac{\bra{\phiL_k}\kket{x_0}}{\mu_k}}
&\text{for FP},
\end{array}\right.
\label{eq:wk_disc_cont}
\end{equation}
for discrete state (DS) and Fokker-Planck (FP) dynamics, respectively, which
must satisfy $\sum_kw_k=1$ with the first nonzero weight being
strictly positive $w_1(x_0)>0$, and where we introduced $\sigma_\pm\equiv\operatorname{sign}(a-x_0)$.
Note that the first line of Eq.~\eqref{eq:wk_disc_cont}, i.e. the DS case,
is equivalent to $w_k(x_0)=\sum_{x\neq a} \bra{x}\kket{\phiR_k} \bra{\phiL_k}\kket{x_0}$
with $\mL_a$ from Eq.~\eqref{eq:La_def}. 
In the case of FP dynamics the second line of
Eq.~\eqref{eq:wk_disc_cont} is equivalent to
 $w_k(x_0)\equiv \int \bra{x}\kket{\phiR_k} \bra{\phiL_k}\kket{x_0}\dd x$,
which follows from a partial integration using both
Eq.~\eqref{eq:FPgen} and Eq.~\eqref{eq:decomp_FP}.

In the case of $x=a$ the renewal theorem \eqref{eq:renewal} has the simple interpretation:
a system being in state $a$ at time $t$ must have arrived at that point
at some earlier time $\tau$ for the first time  ($\tau\le t$),
and then returned to the same position again at time $t$, where $\tau=
t $ corresponds to the time of first arrival.

Laplace transforming the renewal theorem \eqref{eq:renewal}, where a generic function $f$ is transformed according to $\tilde{f}(s)\equiv\int\e^{-s t}f(t)\dd t$, 
yields \cite{sieg51}
\begin{equation}
 \tilde{\wp}_a(s|x_0)=\frac{\tilde{P}(x,s|x_0)}{\tilde{P}(x,s|a)}
 =\frac{\sum_k(s+\lambda_k)^{-1}\bra{x}\kket{\psiR_k}\bra{\psiL_k}\kket{x_0}}{\sum_k(s+\lambda_k)^{-1}\bra{x}\kket{\psiR_k}\bra{\psiL_k}\kket{a}}.
\label{eq:renewal_var}
 \end{equation}
Based on this well known renewal theorem we construct in the following
section a method that allows to determine explicitly the first passage time statistics $\wp_a (t|x_0)$ exactly in terms of the relaxation process, i.e., we
render Eq.~\eqref{eq:renewal_var} explicit in the time domain.

\section{Principal result for discrete state systems}
\label{sec:main_discrete}
Starting from the renewal theorem \eqref{eq:renewal_var}, we now derive an expression for the first passage time density
for discrete state Markov processes
in terms of
relaxation modes in the following three steps.
The first step involves a crucial relation
between the eigenvalues of the
relaxation process $\lambda_k$ and absorption process $\mu_k$, which
are here shown to interlace
\begin{equation}
 \lambda_{k-1}\le \mu_{k}\le \lambda_k
 \label{eq:interlace}
\end{equation}
for $k=1,\ldots,\Nmax$.
For effectively one dimensional finite lattice models
with the target at an outer edge
these inequalities become strict
\begin{equation}
 \lambda_{k-1}<\mu_k<\lambda_k,
 \label{eq:interlace_1d}
\end{equation}
which will also apply identically to Fokker-Planck dynamics discussed in Sec.~\ref{sec:main_cont} in which case we formally assume $\Nmax=\infty$.
In the second step we exactly express the first
passage eigenvalues $\mu_k$ in the form of a Newton series of
determinants of
almost triangular matrices, which generalizes the result for the slowest mode $\mu_1$ from \cite{gode16} to all first passage modes.
The third and final step corresponds to a straightforward application
of the residue theorem, which is used to determine the first passage
weights $w_k(x_0)$.

\subsection{Interlacing of eigenmodes (step 1)}
\label{sec:interlace_proof}

For a discrete system with $\Nmax+1$ states the eigenvalues $\lambda_k$ and $\mu_k$
correspond to the roots of the respective characteristic polynomials
\begin{equation}
 \begin{aligned}
  \chi(s)&\equiv\det(\id s-\mL)=s\prod_{i=1}^\Nmax(s+\lambda_i),\\
  \chi_a(s)&\equiv\det(\id s-\mLa)=s\prod_{i=1}^\Nmax(s+\mu_i),
   \label{eq:chi_defs}
 \end{aligned}
\end{equation}
i.e., $\chi(-\lambda_k)=0$ and $\chi_a(-\mu_k)=0$.
Inserting Eq.~\eqref{eq:La_def}, which is $\mLa=\mL-\mL\ket{a}\bra{a}$, into the
second characteristic polynomial \eqref{eq:chi_defs}
and using the matrix
determinant lemma establishes a link between the two characteristic polynomials
\begin{equation}
 \chi_a(s)=\chi(s)+\bra{a}\adj(\id s-\mL)\mL\ket{a},
 \label{eq:chi_a_pre}
\end{equation}
where $\adj(\matr{A})$ is called the adjugate of a matrix $\matr{A}$ satisfying
Cramer's rule $\matr{A}\adj(\matr{A})=\det(\matr{A})\id$. We note that
the same mathematical concepts have been used  recently to determine
the stalling distribution of irreversibly driven systems
(cf. deletion-contraction formula in \cite{pole17,bisk17}).

The adjugate of a diagonal matrix $\matr{D}$
with elements $D_{ii}=d_i$ ($D_{ij}=0$ if $i\neq j$) is diagonal as well, with elements $\adj(\matr{D})_{ii}=\prod_{j\neq i}d_j$. Consequently, the bi-orthogonal
expansion \eqref{eq:decomp_relax}
implies
\begin{equation}
 \adj(\id s-\mL)=\sum_{i=0}^\Nmax \ket{\psiR_i}\bra{\psiL_i}\prod_{\substack{j=0\\j\neq i}}^\Nmax(s+\lambda_j),
\end{equation}
which inserted into Eq.~\eqref{eq:chi_a_pre}
gives
\begin{align}
 \chi_a(s)&=\chi(s)-\sum_{i=0}^\Nmax\bra{a} \kket{\psiR_i}\bra{\psiL_i}\kket{a}\lambda_i\prod_{\substack{j=0\\j\neq i}}^\Nmax(s+\lambda_j)
 \label{eq:chi_a_pre2},
\end{align}
where we used the eigenvalue equation
$\bra{\psiL_i}\mL\ket{a}=-\lambda_i\bra{\psiL_i}\kket{a}$. Eq.~\eqref{eq:chi_a_pre2}
constitutes an essential step in our calculations, which allows us to
express the diagonal of the relaxation propagator $\tilde{P}(a,s|a)$
solely in terms of eigenvalues $\mu_k$ and $\lambda_k$ (see the following subsection for more details).

Moreover,
the characteristic polynomials of the first passage process $\chi_a(s)$ and relaxation process $\chi(s)$ change
sign one after the other, since detailed balance imposes $\bra{a}\kket{\psiR_i}\bra{\psiL_i}\kket{a}
\ge0$ for all $i=0,\ldots,\Nmax$,
which proves that the eigenvalues of the first passage process $\mu_k$ and eigenvalues of the relaxation process $\lambda_k$
interlace according to Eq.~\eqref{eq:interlace}.
We note that this result is directly related to the interlacing of eigenvalues generated from a  ``lumping'' of
states which is proven in \cite{gron08}.
In the following paragraph we briefly discuss the scenario, in which
the interlacing of eigenvalues becomes strict \eqref{eq:interlace_1d},
which will be the case for systems with
Fokker-Planck dynamics discussed in Sec.~\ref{sec:reflecting_wall}.

\begin{figure}
\centering
\includegraphics{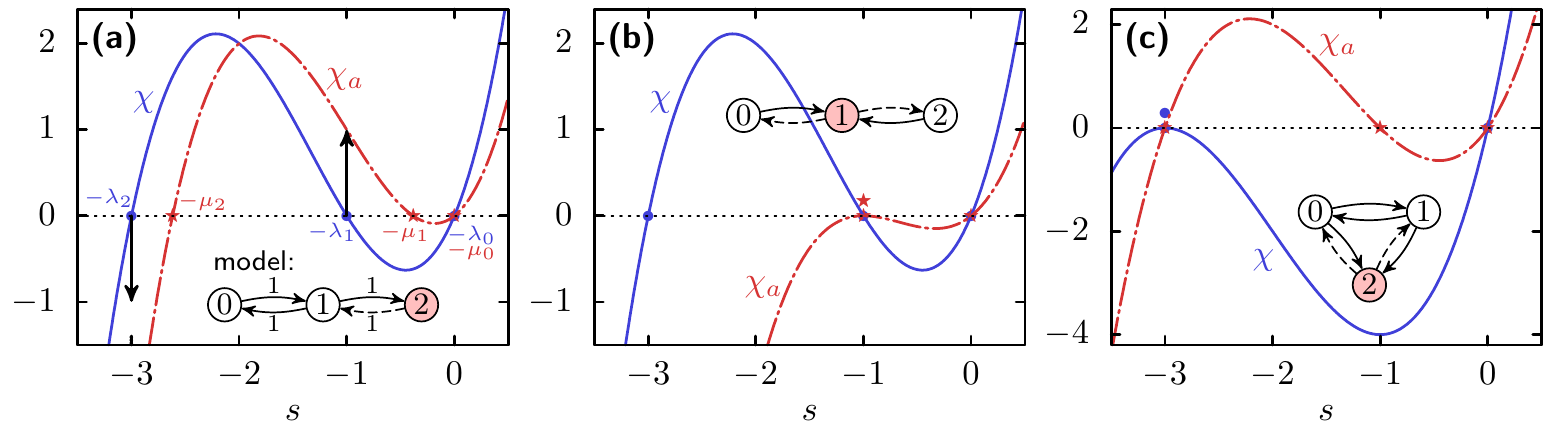}
\caption{Characteristic polynomials $\chi$ (solid blue line) and $\chi_a$ (dash-dotted red line) for simple three state models (see insets) along real axis in $s$.
(a)~Linear chain of states with the absorbing state at the border $a=2$.
$\chi$ crosses the $s$ axis at $s=-3,-1,0$ which
correspond to $-\lambda_2,-\lambda_1,-\lambda_0$, respectively. (b)~The same model as in (a) but with the absorbing state at $a=1$. The first eigenmode $\ket{\psiR_1}=(1,0,1)^\T$ vanishes at the target. (c) Fully connected three state model, in which $\lambda_1=\lambda_2$. All rates in (a)-(c) are set to 1 (relaxation process). The transition rates away from absorbing point (dashed arrows) are set to zero (absorption). We note this special choice of rates
deliberately generates a multiplicity of the first relaxation mode in (b)
and the first passage mode in (c).}
\label{fig:illustration_proof}
\end{figure}

The stronger condition  \eqref{eq:interlace_1d} holds if all eigenfunctions are nonzero at the target $|\bra{a}\kket{\psiR_k}|>0$ and all relaxation eigenvalues are non-degenerate, that is, $\lambda_{i-1}<\lambda_i$ for all $i=1,\ldots,\Nmax$.
One can show that this condition is always trivially satisfied for 1-dimensional models ($L(x,x')=0$ if $|x-x'|>1$), in which the target $a$ is placed at the border (e.g., $a=\Nmax$ or $a=0$); see inset
of Fig.~\mysubref{fig:illustration_proof}{a} for such an exemplary 3-state system.

Inserting the relaxation eigenvalues $s=-\lambda_k$ into the characteristic polynomial
of the first passage process \eqref{eq:chi_a_pre2}
yields
\begin{align}
 \chi_a(-\lambda_k)
 &=(-1)^{k+1}\lambda_k
 \bra{a}\kket{\psiR_i}\bra{\psiL_i}\kket{a}
  \prod_{\substack{i=0\\i\neq k}}^\Nmax |\lambda_i-\lambda_k|,
 \label{eq:oszi_1}
\end{align}
where we used the relations
$\lambda_i-\lambda_k< 0$ for all $i<k$ and
$\lambda_i-\lambda_k>0$ for all $i>k$, as well as $\chi(-\lambda_k)=0$.
Since for $k\ge 1$ each eigenvalue is positive ($\lambda_k>0$),
the characteristic polynomial of the first passage process $\chi_a(-\lambda_k)$
is
equal to $(-1)^{k-1}$ multiplied by a positive constant.
Consequently, 
$\chi_a$ changes sign exactly once between any two consecutive relaxation modes $-\lambda_k<s<-\lambda_{k-1}$.
The fact that $\chi_a$ and $\chi$ are polynomials of the same degree $\Nmax+1$ forbids
more than a single root,
and hence implies the strict interlacing of eigenvalues from
Eq.~\eqref{eq:interlace_1d}, which completes the proof.
The aforementioned reasoning is illustrated in Fig.~\mysubref{fig:illustration_proof}{a} for a simple three state model
in which the vertical arrows represent Eq.~\eqref{eq:oszi_1}.

For fine-tuned systems in which the target is not located at the very outer position
(see e.g., Fig.~\mysubref{fig:illustration_proof}{b}) or systems that are not effectively one dimensional
(see e.g., Fig.~\mysubref{fig:illustration_proof}{c}) the strict interlacing theorem \eqref{eq:interlace_1d} can be violated,
whereas the ``slightly weaker'' interlacing condition \eqref{eq:interlace}  still holds.

\subsection{Diagonal of the relaxation propagator in terms of bare eigenvalues}
\label{subsec:auto}
Using the results from the previous subsection we are now in the position to
represent $\tilde{P}(a,s|a)$ (i.e. Eq.~\eqref{eq:propagator} with $x_0=x=a$),
using only the eigenvalues of both the first passage process and the
relaxation process, $\mu_k$ and $\lambda_k$, respectively.
Laplace transforming the eigenmode expansion in Eq.~\eqref{eq:propagator} assuming $x_0=x=a$ yields
\begin{equation}
 \tilde {P}(a,s|a)
 =\frac{P_\text{eq}(a)}{s}+\sum_{k=1}^\Nmax\frac{\bra{a}\kket{\psiR_k}\bra{\psiL_k}\kket{a}}{s+\lambda_k},
 \label{eq:Paa_eig}
\end{equation}
where we identified the equilibrium probability density $\bra{a}\kket{\psiR_0}\bra{\psiL_0}\kket{a}=P_\text{eq}(a)$ in the first term.
Comparing now $\tilde{P}(a,s|a)$ in \eqref{eq:Paa_eig}
with $\chi_a$ from Eq.~\eqref{eq:chi_a_pre2} and $\chi$ from Eq.~\eqref{eq:chi_defs}
yields after some algebra
\begin{equation}
 \tilde P(a,s|a)=\frac{\chi_a(s)}{s\chi(s)}=\frac{1}{s}\prod_{i=1}^\Nmax \frac{(s+\mu_i)}{(s+\lambda_i)}.
 \label{eq:Paa_prod}
\end{equation}
The second equality in Eq.~\eqref{eq:Paa_prod} follows from Eq.~\eqref{eq:chi_defs}.
Hence, $\tilde{P}(a,s|a)$ encodes
the eigenvalues of both, the relaxation and the first passage processes.
Due to Eq.~\eqref{eq:Paa_eig} $\tilde P(a,s|a)$ contains only simple poles
and decreases monotonically in $s$ between any two consecutive poles
since $\bra{a}\kket{\psiR_k}\bra{\psiL_k}\kket{a}\ge0$.
If $\bra{a}\kket{\psiR_k}\bra{\psiL_k}\kket{a}>0$ (e.g., 1d models
with the target at at the border) each root $\lambda_k$ of $\tilde P(a,s|a)$ represents a first passage eigenvalue $s=-\mu_k$ ($k\ge 1$), which is located in between two relaxation modes
$\lambda_{k-1}<\mu_k<\lambda_k$, thus providing an alternative proof of relation \eqref{eq:interlace_1d} \cite{keil64}.
In the following section we determine the roots of the diagonal of the propagator explicitly, which due to
Eq.~\eqref{eq:Paa_prod} correspond to first passage eigenvalues $\mu_k$.

Let us briefly reformulate $ \tilde P(a,s|a)$ 
in a way that  can also be applied to continuous systems with an infinite number of states.
Isolating the equilibrium probability, which is the first term in Eq.~\eqref{eq:Paa_eig}, from the product formula \eqref{eq:Paa_prod}
yields
\begin{equation}
  \tilde {P}(a,s|a)=\frac{P_\text{eq}(a)}{s}\prod_{k=1}^\Nmax\frac{(1+s/\mu_k)}{(1+s/\lambda_k)}.
  \label{eq:Paa_mulambda}
\end{equation}
Since $\mu_k,\lambda_k$ increase monotonically with $k$ we will later be able to
adopt these results to systems governed by Fokker-Planck dynamics, which formally
corresponds to the limit $\Nmax\to\infty$ for which the product in
Eq.~\eqref{eq:Paa_mulambda} still
converges.

\subsection{From the relaxation spectrum to the first passage time spectrum (steps 2 and 3)}
\label{sec:relax_to_FP_disc}
Based on the interlacing theorem presented in Eq.~\eqref{eq:interlace}, which is
also given in Eq.~(4) in a related work \cite{hart18_arxiv}, we can determine the full first passage time spectrum
$\{\mu_k,w_k(x_0)\}$ from the corresponding relaxation spectrum,
$\{\lambda_k,\ket{\psiR_k},\bra{\psiL_k}\}$.
For simplicity we first consider the eigenvalues to be both ordered
$\lambda_k<\lambda_{k+1}$ and non-degenerate, and also assume that $\bra{a}\kket{\psiR_k}\bra{\psiL_k}\kket{a}>0$ holds for all values of $k$.
The extension to situations with $\bra{a}\kket{\psiR_k}\bra{\psiL_k}\kket{a}=0$, which also includes degenerate eigenvalues, for some $k$ is straightforward and will be dealt with at the end of this subsection.

Before determining the weights $w_k$, we first determine the first passage
eigenvalues $\mu_k$, which were shown to be encoded in the roots of $\tilde P(a,s|a)$ in Eq.~\eqref{eq:Paa_prod}.
We introduce the $k^*$th ``modified diagonal of the propagator''
\begin{align}
 F_{k^*}(s)&
 \equiv (s+\lambda_{k^*})\tilde P(a,s|a)\nonumber\\
 &=\bra{a}\kket{\psiR_{k^*}}\bra{\psiL_{k^*}}\kket{a}+(s+\lambda_{k^*})\sum_{ \hidewidth\substack{l=0\\l\neq k*}\hidewidth}^\Nmax\frac{\bra{a}\kket{\psiR_{l}}\bra{\psiL_{l}}\kket{a}}{s+\lambda_i},
 \label{eq:Fk_def}
\end{align}
which still encodes all of the first passage eigenvalues $\{\mu_k\}$
according to Eq.~\eqref{eq:Paa_prod}, i.e., it has exactly the same roots
as $\tilde P(a,s|a)$. However, in contrast to $\tilde P(a,s|a)$ the modified function $F_{k^*}(s)$ is strictly concave within the interval $-\lambda_{k^*+1}< s<-\lambda_{k^*-1}$,
which can easily be confirmed by taking the second derivative and realizing that $\ddot F_{k^*}(s)\equiv\del_s^2F_{k^*}(s)<0$ holds within the region of interest $-\lambda_{k^*+1}< s<-\lambda_{k^*-1}$.

For $k^*=k$ and $k^*=k-1$ the modified functions   $F_{k}(s)$ and $F_{k-1}(s)$ both are strictly concave within the interval $-\lambda_k<s<-\lambda_{k-1}$
and, consequently, also locally concave around the $k$th first passage eigenvalue $s=-\mu_k$, i.e., $\ddot F_{k}(-\mu_k)$ and $\ddot F_{k-1}(-\mu_k)<0$.
Moreover, both functions $F_{k}(s)$ and $F_{k-1}(s)$ allow a Taylor
expansion around the  midpoint $\bar\mu_k\equiv(\lambda_k+\lambda_{k-1})/2$
that converges within the whole interval $-\lambda_k<s<-\lambda_{k-1}$
including the root $s=-\mu_k$ at which $F_{k}(-\mu_k)=F_{k-1}(-\mu_k)=0$.

The method we present in the following is an analytical technique based on the principles of Newton iteration, which is a
simple root finding algorithm that is guaranteed to work for functions that are both negative and concave between the starting point and the first root.
Hence, to determine the $k$th eigenvalue we accordingly choose the
modified function 
\begin{equation}
 f(s,k)= F_{k^*}(s),
\label{eq:fsk}
\end{equation}
such that
\begin{equation}
 k^*=
 \left\{\begin{array}{ll}
  k&\text{if $F_k(-\bar\mu_k)<0$,}\\
  k-1&\text{otherwise}.
 \end{array}\right.
 \label{eq:k*},
\end{equation}
which guarantees both negativity $f(s,k)\le 0$ and concavity $\del_s^2f(s,k)\le0$ between
$s=-\bar\mu_{k}$ and $s=-\mu_k$.

According to the interlacing theorem \eqref{eq:interlace_1d} $s=-\mu_k$ is the only zero $f(-\mu_k,k)=0$ within the interval $-\lambda_k<s<-\lambda_{k-1}$.
With the midpoint starting condition $\bar\mu_k=(\lambda_{k}+\lambda_{k-1})/2$
the $k$th first passage eigenvalue can be represented exactly in
a series of determinants of almost triangular matrices
\begin{equation}
 \mu_k=\bar\mu_k+\sum_{n=1}^\infty \frac{f_0(k)^n}{f_1(k)^{2n-1}}\frac{\det\mcA_n(k)}{(n-1)!},
 \label{eq:mu_k_newton}
\end{equation}
where $f_n(k)$ is the $n$th derivative of $f(s,k)$ as defined in \eqref{eq:fsk} with respect to $s$ at $s=-\bar\mu_k$, and
$\mcA_n(k)$ stands for an almost triangular matrix 
with elements \cite{gode16}
\begin{equation}
 \cA_n^{i,j}(k)=\frac{f_{i-j+2}(k)\Theta(i-j+1)}{(i-j+2)!}
 \Big[n(i-j+1)\Theta(j-2)
 +i\Theta(1-j)+j-1\Big],
 \label{eq_def_An}
\end{equation}
with $\Theta(l)$ denoting the Heaviside step function ($\Theta(l)=1$ if $l\ge0$)
and $i,j=1,2,\ldots,n-1$. Moreover, we adopt the convention $\det\mcA_1(k)=1$.
We note that this method generalizes 
the method recently derived to determine the slowest first passage mode $\mu_1$ \cite{gode16} to all first passage eigenmodes $\mu_k$.

Let us briefly repeat the two crucial steps towards Eq.~\eqref{eq:mu_k_newton}. First, the interlacing theorem \eqref{eq:interlace_1d} guarantees that the Taylor series 
$f(s,k)=\sum_i f_i(k)(s+\bar\mu_k)^i$ around the midpoint
$\bar\mu_k=(\lambda_{k}+\lambda_{k-1})/2$ converges in the entire spectral interval 
$-\lambda_{k}<s<-\lambda_{k-1}$,
which also contains the first passage eigenvalue $s=-\mu_k$. Second,
due to $F_{k^*}(s)$ in
Eqs.~\eqref{eq:Fk_def}-\eqref{eq:k*}
the function $f(s,k)$ is strictly concave and negative between $s=-\bar\mu_k$ and $s=-\mu_k$, which in turn guarantees the convergence of the
explicit Newton series \eqref{eq:mu_k_newton}.

Eqs.~\eqref{eq:Fk_def}-\eqref{eq:k*} provide a universal method for determining
explicitly first passage eigenvalues from the corresponding relaxation spectrum
and constitute the central result of this work. We show in the \ref{sec:A:mu1} a simpler derivation of $\mu_1$ as well as a compact approximation of the principal first passage eigenvalue $\mu_1$, which is particularly useful in the case of time scale separation $\mu_1\ll\lambda_1$ (or $\lambda_1\ll\lambda_2$).
Furthermore, \ref{sec:A:mu1} provides a generalization of the long time
asymptotics from systems with reversible dynamics to irreversibly
driven systems.

In the following we briefly comment on the practical
implementation of the exact result for $\mu_k$  to render Eqs.~\eqref{eq:Fk_def}-\eqref{eq:mu_k_newton} fully explicit. The weights $w_k$ will be determined afterwards in this subsection.
The $n$th derivative of $F_{k^*}(s)$ with respect to $s$
at $s=-\bar{\mu}_k$,  $f_n(k)\equiv\partial_s^nf(s,k)|_{s=-\bar{\mu}_k}$,
can be written explicitly as
\begin{equation}
\begin{aligned}
&f_0(k)=\bra{a}\kket{\psiR_{k^*}}\bra{\psiL_{k^*}}\kket{a}+\sum_{\hidewidth l|l\neq k^*\hidewidth}\bra{a}\kket{\psiR_{l}}\bra{\psiL_{l}}\kket{a}\frac{(\bar\mu_k-\lambda_{k^*})}{(\bar\mu_k-\lambda_l)},\\
&f_{n\ge 1}(k)=n!\sum_{\hidewidth l|l\neq k^*\hidewidth}\bra{a}\kket{\psiR_{l}}\bra{\psiL_{l}}\kket{a}\frac{(\lambda_l-\lambda_{k^*})}{(\bar\mu_k-\lambda_l)^{n+1}},
\label{eq:fn}
\end{aligned}
\end{equation}
where $k^*=k$ or $k^*=k-1$ is chosen according to Eq.~\eqref{eq:k*}. Note that condition \eqref{eq:k*} is equivalent
to the condition $f_0(k)\le 0$, implying the first line of Eq.~\eqref{eq:fn}
to be either negative for $k^*=k$ or for $k^*=k-1$,
i.e. one has to evaluate the first line of Eq.~\eqref{eq:fn} for $k^*=k$:
if $f_0(k)>0$ one must to change $k^*$ to $k^*=k-1$ and reevaluate $f_0(k)$.
Once one has determined $k^*(k)$ and $f_0(k)$ one can proceed with the second line of Eq.~\eqref{eq:fn} to determine $f_n(k)/n!$ and insert the result in the almost triangular matrix \eqref{eq_def_An}.
The determinant of almost triangular matrices can be
calculated elegantly using the simple recursion relation from \cite{cahi02}, see also \cite{jia18} for an efficient numerical implementation.

Having obtained the first passage eigenvalues,
the weights of the first passage time distribution can be calculated
using the standard residue theorem. The Laplace transform of the spectral expansion of the first passage time density \eqref{eq:FPTdensity_def} reads
\begin{equation}
 \tilde{\wp}_a(s|x_0)\equiv\sum_k \frac{w_k(x_0)\mu_k}{s+\mu_k}.
\end{equation}
Using the residue theorem to invert the Laplace transformed renewal theorem
\eqref{eq:renewal_var}
yields
\begin{align}
 w_k(x_0)&=\frac{\tilde P(a,-\mu_k|x_0)}{\mu_k\dot{\tilde P}(a,-\mu_k|a)}
 \nonumber\\
&=\frac{\sum_l(1-\lambda_l/\mu_k)^{-1}\bra{a}\kket{\psiR_l}\bra{\psiL_l}\kket{x_0}}{\sum_l(1-\lambda_l/\mu_k)^{-2}\bra{a}\kket{\psiR_l}\bra{\psiL_l}\kket{a}},
\label{eq:wk_explicit}
 \end{align}
where $\dot{\tilde{P}}(a,s|a)=\del_s\tilde{P}(a,s|a)$ is taken at $s=-\mu_k$. The explicit Newton series \eqref{eq:mu_k_newton} along with the first passage weights \eqref{eq:wk_explicit} fully characterize the first passage time
distribution $\wp_a(t|x_0)=\sum_{k} w_k(x_0)\mu_k\e^{-\mu_k t}$ in terms of relaxation
eigenmodes $\{\lambda_k,\psiR_k\}$. This completes our third and final step, which allows, for the first time, to analytically deduce first passage time statistics directly from relaxation eigenmodes. We call this relation the explicit \emph{forward duality} between first passage and relaxation. This completes the central result of this paper.

The spectral representation is very useful for determining the moments
of the first passage time, $\langle t^n\rangle\equiv\int
t^n\wp_a(t|x_0)\dd t=n!\sum_{k}w_k(x_0)\mu_k^{-n}$.
Moreover, as explained in more detail in a related work \cite{hart18_arxiv}, the
full spectral expansion is required for a correct explanation of
kinetics in the 
so-called few encounter limit, where $N$ molecules
starting from position $x_0$ are searching for the target at $a$. 
The probability density that the first molecule out of $N$ arrives at
time $t$ at $a$ for the first time for this case
becomes $\wp_a^{(N)}(t|x_0)=N\wp_a(t|x_0)[\int_t^\infty \wp_a(\tau|x_0)]^{N-1}\dd \tau$, which can be understood as follows.
The probability that the first $N-1$ molecule have not yet reached the target will be given by $[\int_t^\infty \wp_a(\tau|x_0)\dd \tau]^{N-1}$, while the $N$th particle
arrives at $a$ with a rate $\wp_a(t|x_0)$; hence the probability density that any particle
out of $N$ molecules arrives at the target for the first time
according to $\wp_a^{(N)}(t|x_0)$. Further details of the $N$-particle
problem and in particular the physical implications of the
few-encounter limit are discussed in a related study
\cite{hart18_arxiv}.

Let us now briefly generalize the method to systems with degenerate eigenvalues
or vanishing relaxation modes.
An eigenfunction that vanishes at the target $\bra{a}\kket{\psiR_k}=0$
will have a vanishing spectral weight as a result of
Eq.~\eqref{eq:wk_explicit}. Hence, `manually' removing such modes will not
affect the first passage time distribution $\wp_a$. Moreover, if a relaxation eigenvalue $\lambda_k$ 
is degenerate we define 
\begin{equation}
 \Psi_{k}(a,x_0)\equiv\sum_{k'|\lambda_{k'}=\lambda_{k}}\bra{a}\kket{\psiR_{k'}}\bra{\psiL_{k'}}\kket{x_0},
\end{equation}
and replace $\bra{a}\kket{\psiR_k}\bra{\psiL_k}\kket{x_0}\to\Psi_{k}(a,x_0)$
as well as $\bra{a}\kket{\psiR_k}\bra{\psiL_k}\kket{a}\to\Psi_{k}(a,a)$
and take the sums in Eq.~\eqref{eq:renewal_var} over all different values of $\lambda_k$.
After renumbering all distinct contributing eigenvalues we obtain a strict
interlacing \eqref{eq:interlace_1d}. Therefore, we can apply our standard forward duality also to degenerate eigensystems.
In the next subsection we will briefly derive a formal \emph{backward duality}
after which we reformulate the results from this subsection to
continuous Fokker-Planck dynamics.

\subsection{Backward duality}
\label{sec:inverse_duality}
In contrast to the explicit \emph{forward duality}, which was presented in the previous subsection, an explicit reverse relation in the time-domain could not be established.
In Laplace space, however, the forward duality can be inverted to give
a backward duality as follows. Inserting the first passage generator
$\mLa=\mL-\mL\ket{a}\bra{a}$ from \eqref{eq:La_def} into the Laplace transform
of the propagator $\tilde{P}(a,s|x_0)=\bra{a}(\id s-\mL)^{-1}\ket{x_0}$ and
using the Sherman-Morrison-Woodbury formula
yields
\begin{equation}
 \tilde P(a,s|x_0)=\frac{\bra{a}(\id s-\mLa)^{-1}\ket{x_0}}{1-\bra{a}(\id s-\mLa)^{-1}\mL\ket{a}}.
 \label{eq:Dual_reverse_1}
\end{equation}
Let us now insert the expression for the first passage time distribution from Eq.~\eqref{eq:pFP_def}, which can be written as $\wp_a(s|x_0)=s\bra{a}(\id s-\mLa)^{-1}\ket{x_0}$, 
into Eq.~\eqref{eq:Dual_reverse_1} to obtain
\begin{equation}
\tilde{P}(a,s|x_0)=\frac{ \tilde{\wp}_a(s|x_0)/s}{1 -\sum_{x=0}^\Nmax L(x,a)\tilde{\wp}_a(s|x)/s},
\label{eq:inverse_duality}
\end{equation}
where $L(x,a)=\bra{x}\mL\ket{a}$ is the generator of the relaxation process.
Notably, this is expression corresponds to the \emph{backward duality}
and is the formal inverse of the renewal theorem,
where $\tilde\wp_a(s|x_0)/s$ is the Laplace transform of the cumulative first passage time distribution $\int_0^t\wp_a(\tau|x_0)\dd \tau$.

\section{Principal result for Fokker-Planck dynamics}
\label{sec:main_cont}
\subsection{Greens function with natural boundaries}
We restrict our discussion to effectively 1-dimensional dynamics, which include
diffusion in $d$ dimensions in an isotropic potential as discussed in \cite{gode16}, where $d$ may also be fractal.
Introducing an absorbing target at position $a$ splits the
first passage problem into two cases (I) $x_0<a$ and (II) $x_0>a$.
Case (I) corresponds to an absorption from the left, and case (II) to an absorption from the right.
In the following paragraph we demonstrate that all first passage modes $\mu_k$
of both distinct cases (I) and (II) are entirely encoded in  $\tilde P(a,s|a)$, which allows to formulate the results from Sec.~\ref{sec:relax_to_FP_disc}
also for systems with Fokker-Planck dynamics.

Laplace transforming the Fokker-Planck equation \eqref{eq:FPgen} yields
\begin{equation}
 (\mL-s)\tilde P(x,s|x_0)=-\delta(x-x_0)
 \label{eq:FP_greens}
\end{equation}
where $\mL=-\del_xD[\beta U'(x)+\del_x]$ and $x_0$ is the initial position of the relaxation process.
Eq.~\eqref{eq:FP_greens} is a inhomogeneous linear differential equation
which can be solved using the standard Green's function approach.
First, we find the two independent solutions $v_\pm(x,s)$ of the homogeneous problem $(\mL-s)v_\pm(x,s)=0$, where we use the label ``$-$'' and ``$+$''
for the solution satisfying the left and right boundary condition, respectively.
That is, a diffusion process within an interval $b_-<x<b_+$
imposes the probability current $j_\pm(x,s)\equiv - D[\beta U'(x)+\del_x]v_{\pm}(x,s)$
to vanish at the boundaries, i.e. $j_\pm(b_\pm,s)=0$. The special case of so-called natural boundary conditions correspond to the limit  $\lim_{x\to\pm\infty}v_\pm(x,s)= 0$
or analogously $\lim_{x\to\pm\infty}j_\pm(x,s)= 0$, that is, $b_\pm=\pm\infty$.
The full solution $\tilde P(x,s|x_0)$ of \eqref{eq:FP_greens} is a continuous function in $x$ with a discontinuity of its first derivative at $x=x_0$.
Using the scaled Wronskian
\footnote{For convenience we defined with the scaled Wronskian with the current function $j_\pm(x,s)$ instead of the first derviative $\del_xv_\pm(x,s)$, i.e., $W_s(x)/D$ would represent the standard definition of the Wronskian.}
\begin{align}
W_s(x)
&\equiv D[v_-(x,s)\del_x v_+(x,s)-v_+(x,s)\del_x v_-(x,s)],\nonumber\\
&=v_+(x,s)j_-(x,s)-v_-(x,s)j_+(x,s)\nonumber\\
&= \det 
\begin{pmatrix}
 v_+(x,s)&v_-(x,s)\\
  j_+(x,s)&j_-(x,s)
\end{pmatrix}
\label{eq:Wronskian}
\end{align}
the propagator, which satisfies the proper jump condition of the first derivative (current function) at $x=x_0$, becomes
\begin{equation}
 \tilde P(x,s|x_0)=
 \left\{\begin{array}{ll}
 \displaystyle{\frac{v_+(x,s)v_-(x_0,s)}{W_s(x_0)}} &\text{if $x_0\le x$},\\
  \displaystyle{\frac{v_-(x,s)v_+(x_0,s)}{W_s(x_0)}} &\text{if $x_0\ge x$}.
 \end{array}\right.
 \label{eq:propagator_solution}
\end{equation}
We note that the Wronskian \eqref{eq:Wronskian} is proportional to the Boltzmann factor (see, e.g., Ref.~\cite{keil64}), i.e., $W_s(x)=W_s(x_0)\exp[\beta U(x_0)-\beta U(x)]$. Hence using the renewal theorem \eqref{eq:renewal_var}
and $\tilde P(a,s|a)$ as well as $\tilde P(a,s|x_0)$
from Eq.~\eqref{eq:propagator_solution}
yields
the Laplace transform of the first passage time distribution
\begin{equation}
 \wp_a(s|x_0)=
 \e^{\beta U(x_0)-\beta U(a)}\times
 \left\{\begin{array}{ll}
\displaystyle{\frac{v_-(x_0,s)}{v_-(a,s)}}&\text{if $x_0<a$},\\
\displaystyle{\frac{v_+(x_0,s)}{v_+(a,s)}}&\text{if $x_0>a$}.
 \end{array}\right.
 \label{eq:FPDv+-}
\end{equation}
The two independent functions $v_\pm (x,s)$ are entire functions
without any poles in $s$ \cite{titc62}, and in turn encode in their
roots all first passage eigenvalues $s=-\mu_k$. In particular $v_-$ encodes all first passage modes
from case (I) $x_0<a$, and $v_+$ encodes all first passage modes from case (II), in which the particle is absorbed from the right $x_0>a$. Due to 
Eq.~\eqref{eq:propagator_solution} the zeros of $\tilde P(a,s|a)$ at $s=-\mu_k$ determine the first passage spectrum.
Hence, all results from Sec.~\ref{sec:relax_to_FP_disc} hold identically for continuous systems as well. However, the sums are here not finite, i.e., $\Nmax=\infty$. For example, $\tilde P(a,s|a)$ becomes
\begin{equation}
\tilde P(a,s|a)=\sum_{l=0}^\infty\frac{\psiL_l(a)\psiR_l(a)}{s+\lambda_l},
\label{eq:Paa_FP}
\end{equation}
where $\psiR_l$ is the $l$th right eigenfunction of the Fokker-Planck operator
satisfying $\mL\psiR_l(x)=-\lambda_l\psiR_l(x)$ with the corresponding
left eigenfunction $\psiL_l(a)\propto\e^{\beta U(a)}\psiR_l(a)$ and normalization $\int_{b_-}^{b_+} \psiR_l(x)\psiL_l(x)\dd x=1$.
The first passage modes $\mu_k$ can then be determined with Eqs.~\eqref{eq:mu_k_newton}-\eqref{eq:fn}, where $k^*$ ($k^*=k$ or $k^*=k-1$) must be chosen such that
$f_0(k)<0$  holds in Eq.~\eqref{eq:fn} with $\bra{a}\kket{\psiR_l}\bra{\psiL_l}\kket{a}\equiv\psiL_l(a)\psiR_l(a)$.
Concurrently, the first passage weights follow from Eq.~\eqref{eq:wk_explicit}.

The formal backward duality from Sec.~\ref{sec:inverse_duality},
however, must be adopted as follows. After some tedious algebra we
obtain formally the exact inverse duality in the form of
\begin{equation}
 \tilde P(x,s|x_0)=\sigma_\pm\frac{\e^{\beta U(x_0)-\beta U(x)}\wp_{x_0}(s|x)}{D\frac{\del}{\del x_0}\ln[\wp_{x_0}(s|x)\wp_{x}(s|x_0)]},
 \label{eq:inverse_duality_FP}
\end{equation}
where sign $\sigma_\pm=-1$ if $x_0< x$ and $\sigma_\pm=+1$ if $x_0> x$;
Eq.~\eqref{eq:inverse_duality_FP}
can easily be verified by inserting
the  Wronskian \eqref{eq:Wronskian} and the first passage time distribution \eqref{eq:FPDv+-} into the right hand side
of Eq.~\eqref{eq:inverse_duality_FP}, and comparing the result with the propagator from Eq.~\eqref{eq:propagator_solution}. Notably,  this inverse duality is the continuous version of Eq.~\eqref{eq:inverse_duality}.

\subsection{Relaxation under reflecting boundary conditions and strict
  spectral interlacing}
\label{sec:reflecting_wall}
In the previous subsection the target $a$ divided the phase space into two regions,
which implies that the first passage modes for the cases (I) and (II) separate
into ``left'' and ``right'' modes as well. For example, if $x_1<a$ and $x_2>a$
one of the first passage weights $w_k(x_1)$ or $w_k(x_2)$ must typically be zero for all values of $k$. 
If one uses just the first $M$ modes to approximate the propagator [cf. Eq.~\eqref{eq:Paa_FP}]
the zeros of the right hand side of
\begin{equation}
\tilde P^M(a,s|a)\equiv\sum_{k=0}^{M}\frac{\psiR_k(a)\psiL_k(a)}{s+\lambda_k},
\label{eq:Pxx0_FP_M}
\end{equation}
become approximations of the first passage modes and, hence, the weights
$w_k^M(x_1)$ and $w_k^M(x_2)$ deduced from Eq.~\eqref{eq:Pxx0_FP_M} will only satisfy $w_k^M(x_1)\ll w_k^M(x_2)$
(or $w_k^M(x_1)\gg w_k^M(x_2)$) for finite $M$, i.e., modes from case (I) and (II)
mix. Such a mixing can be avoided entirely if the relaxation process is analyzed with
a reflecting boundary at the target position $a$ ($b_+=a$ or $b_-=a$).

The result for reflecting boundary conditions $j_+(a)=0$ (case (I)) and $j_-(a)=0$  (case (II)) 
is automatically obtained by the following replacement:
\begin{equation}
\begin{aligned}
  v_\pm(x)&\to v_\pm(x,s)j_\mp(a,s)-j_\pm(a,s)v_\mp(x,s),\\
  j_\pm(x)&\to j_\pm(x,s)j_\mp(a,s)-j_\pm(a,s)j_\mp(x,s),
\end{aligned}
\end{equation}
respectively, which inserted into the scaled Wronskian \eqref{eq:Wronskian}
at $x=a$
yields
\begin{equation}
W_s(a)=
\left\{\begin{array}{ll}
v_+(a,s)j_-(a,s) &\text{if  case (I) $x_0<a$},\\
-v_-(a,s)j_+(a,s) &\text{if  case (II) $x_0>a$}.
\end{array}\right.
\end{equation}
Utilizing the Wronskian for the reflecting boundary condition in Eq.~\eqref{eq:propagator_solution} 
yields the diagonal of the propagator in the form
\begin{equation}
 \tilde{P}(a,s|a)\equiv \lim_{\epsilon\to0} \tilde{P}(a\pm\epsilon,s|a)=\pm\frac{v_\pm(a,s)}{j_\pm(a,s)}
 \label{eq:Paa_reflect}
\end{equation}
For two linearly independent functions $v_\pm$ with nonzero Wronskian \eqref{eq:Wronskian} the zeros of $v_\pm$ and $j_\pm$ are different. Hence, the zeros of $v_-$ in $s$ contain
only the first passage modes for the case (I) $x_0<a$, whereas the
zeros of $v_-$ do not contain zeros of first
passage modes corresponding to the case (II) $x_0>a$.

Let us from now on just focus on a case (I), in which $x_0<a$, since case (II) follows by analogy.
For case (I) we consider the Fokker-Planck operator $\mL$ from Eq.~\eqref{eq:FPgen}
with zero current condition at $x=a$ and natural boundary condition for $x\to-\infty$.
To that end we first determine the relaxation eigenvalues
$\lambda_k$ and eigenmodes $\psiR_k{},\psiL_k{}$.
Note that we consider the eigensystem in the presence of a
reflecting wall. Without loss of generality we here explicitly treat
only the ``left''
problem $-\infty<x\le a$, since the opposite ``right'' problem (denoted
later on with $\dagger$) follows
by analogy.  As before we have the normalization $\int_{-\infty}^a\psiL_{k}{}(x)\psiR_{l}{}(x)\dd x=\delta_{kl}$.
Using $\{\psiL_{k}{}(x),\psiR_{l}{}(x)\}$ we now determine the first passage eigenvalues $\mu_k$ as explained in the previous subsection.
The resulting first passage eigenvalues $\mu_k$ will
automatically contain only
first passage modes corresponding to the ``left'' problem. This procedure remarkably simplifies the numerical
determination of the first passage distribution, especially of those
modes that are faster than the slowest mode of the ``right'' problem
(i.e. absorption from the right), $\mu_k>\mu_1^\dagger$, since a small number of modes $M$
in Eq.~\eqref{eq:Pxx0_FP_M} might otherwise be confused with `fantom' modes
from the opposite case $\dagger$. In a related work \cite{hart18_arxiv} we
investigated the ``left'' first passage problem (case (I)) for a triple well potential in the presence of a reflecting
boundary, and found an excellent agreement between the analytical first
passage time distribution and computer simulations extending 
over many orders of magnitude in time using merely
$M=40$ relaxation modes.
Finally, we have to point out that solving an eigenvalue problem $\{\lambda_k,\psiR_k\}$ with reflecting boundary condition is numerically easier than without reflecting boundary, i.e., natural boundaries albeit theoretically easier are numerically harder.

In the following section we apply these theoretical results to a discrete-state protein folding model and for the Ornstein-Uhlenbeck process.

\section{Examples}
\label{sec:examples}

\subsection{Discrete protein folding model}

\begin{figure}
  \centering
\includegraphics{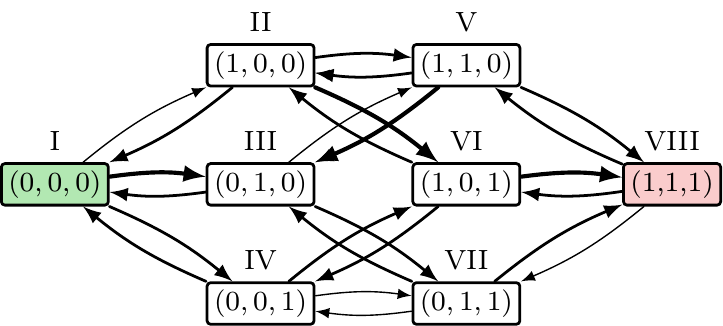}
  \caption{Discrete state protein folding model.
  Each arrow indicates a transition $x\to x'$ with rate $L(x',x)=\exp(F_x-B_{xx'})$, where $B_{xx'}=B_{x'x}$ is the energy barrier between the pair of states
  $x$ and $x'$ and $F_x$ is the free energy of state $x$, where
  $x,x'=\mathrm{I},\ldots,\mathrm{VIII}$. We randomly generated
  100 folding landscapes with $B_{xx'}$ and $F_x$ uniformly distributed within the interval $0\le B_{xx'},F_x\le 4$.
  The results for one particular realization of the landscape are presented in 
  Fig.~\ref{fig:folding_result1}.
  }
  \label{fig:discrete_folding}
\end{figure}

We consider a simple continuous-time Markov state model 
for a protein with three structural elements as shown in
Fig.~\ref{fig:discrete_folding}. The protein starts from an initially unfolded state $x_0=(0,0,0)\equiv\mathrm{I}$,
from which it is searching for the native state $a=(1,1,1)$ through
intermediate states II-VII (see e.g., \cite{prin11,bowm14}).
Each arrow in Fig.~\ref{fig:discrete_folding} indicates a possible transition $x\to x'$ ($x,x'=\mathrm{I},\ldots,\mathrm{VIII}$, $x\neq x'$)
that occurs with a Arrhenius type rate $L(x',x)\equiv\exp(F_x-B_{xx'})$, where
$F_x$ denotes the free energy of state $x$ and $B_{xx'}=B_{x'x}$ the
energy barrier along the transition link $x\leftrightarrow x'$.
The resulting transition matrix satisfies detailed balance $\ln [L(x',x)/L(x,x')]=F_x-F_{x'}$ for all values of $\{F_x,B_{xx'}\}$,
and naturally has negative diagonal elements
$L(x,x)=-\sum_{x'\neq x}L(x',x)$.

To test the power of the method from Sec.~\ref{sec:relax_to_FP_disc}
we set up the $8\times8$ transition matrix
$\mL$ with elements $\bra{x'}\mL\ket{x}=L(x',x)$ for a given set of energy barriers $B_{xx'}$ and free energies $F_x$.
Then we carry out the eigendecomposition of $\mL$, for which we first determine
the eigenvalues $0<\lambda_1,\ldots, \lambda_7$ (with $\lambda_0=0$)
corresponding to the zeros of the characteristic function
\eqref{eq:chi_defs}, $\chi(-\lambda_k)=0$. We then determine the right eigenvectors $\ket{\psiR_k}$
by solving $\mL\ket{\psiR_k}=-\lambda_k\ket{\psiR_k}$ for $k=0,\ldots,7$.
The corresponding left eigenvectors, which solve $\bra{\psiL_k}\mL=-\bra{\psiL_k}\lambda_k$,
have components
$\bra{\psiL_k}\kket{x}=\mathcal{N}^{-1}\e^{F_x}\bra{x}\kket{\psiR_k}$,
where $\mathcal{N}=\sum_{x=\mathrm{I}}^{\mathrm{VIII}}\e^{F_x}|\bra{\psiR_k}\kket{x}|^2$ is a normalization factor. We take the function
$F_{k^*}$ as defined in Eq.~\eqref{eq:Fk_def}, where
$\Psi_k=\bra{a}\kket{\psiR_k}\bra{\psiL_k}\kket{a}$ and
$\bar{\mu}_k=(\lambda_{k}+\lambda_{k-1})/2$ with $k=0,\ldots,7$, and choose $k^*(k)=k,k-1$ according to Eq.~\eqref{eq:k*}, which guarantees $f(s,k)=F_{k^*(k)}(s)$
to be negative at $s=-\bar \mu_k$.
The truncated Newton series \eqref{eq:mu_k_newton} involving  the first $N$ terms
is then given by
\begin{equation}
 \mu_k^N=\bar\mu_k+\sum_{n=1}^N \frac{f_0(k)^n}{f_1(k)^{2n-1}}\frac{\det\mcA_n(k)}{(n-1)!},
 \label{eq:mu_k_newton_N}
\end{equation}
where $\det\mcA_n(k)$
is the determinant of the almost triangular matrix from Eq.~\eqref{eq_def_An}
and 
$f_i(k)$ is the $i$th derivative of $f(s,k)$ at $s=-\mu$,
with explicit formulas given in Eq.~\eqref{eq:fn}.
The weights $w_k(x_0)$ are determined using Eq.~\eqref{eq:wk_explicit},
i.e. by inserting $\mu_k\to\mu_k^N$.
The calculations are performed for 100 randomly generated folding landscapes chosen 
as described in the caption to Fig.~\ref{fig:discrete_folding}.
\begin{figure}
 \includegraphics{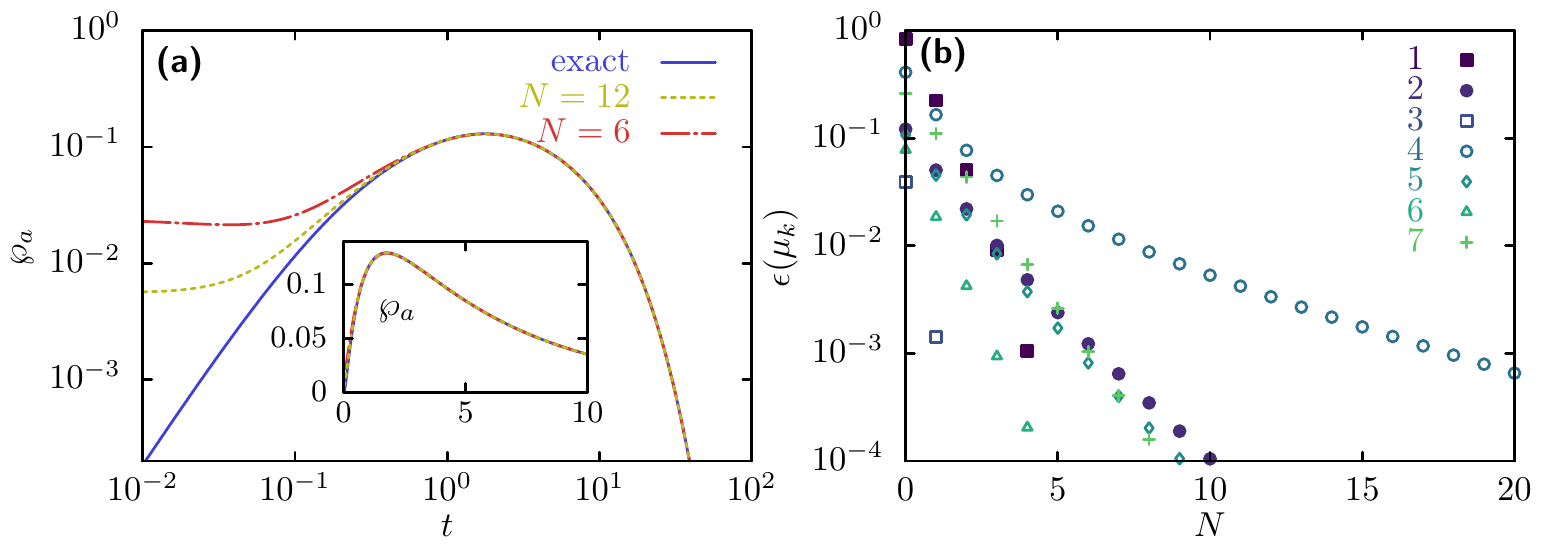}
 \caption{Results for a particular realization of the folding landscape. (a) First passage time distribution as
   function of time $t$. The Inset depicts the same data but on a
   linear time scale. (b) Relative error of the first passage eigenvalue $\epsilon(\mu_k)\equiv |\mu_{k,N}-\mu_k|/\mu_k$, where
 $\mu_{k,N}$ is the finite version of \eqref{eq:mu_k_newton}, where
$n=1,\ldots,N$.}
\label{fig:folding_result1}
\end{figure}
In Fig.~\ref{fig:folding_result1} we present the results for one particular realization of the folding landscape.
Fig.~\mysubref{fig:folding_result1}{a} displays the first passage time distribution
for $N=6$ and $N=12$ on a doubly-logarithmic scale. The solid line
represents the first passage time distribution obtained via a
numerical diagonalization of $\mL_a$. The corresponding duality solutions nicely overlap with the numerical result even on relatively short time scales (see inset for a plot with linear scales).

Having obtained the full distribution of first passage times is
important for understanding kinetics in the so-called few encounter
limit \cite{gode16}, in which for example 100 molecules are
simultaneously searching for a state $a$. This scenario is indeed
biologically relevant, for example, in the misfolding-triggered protein
aggregation, which in turn leads to numerous diseases (see
\cite{hart18_arxiv} for a more detailed discussion). Namely, as soon as the first protein
molecule spontaneously misfolds it creates a nucleation site for further
downhill misfolding and aggregation events ultimately leading to a
macroscopic insoluble toxic aggregate. 

In such a scenario the 
typical timescale of first arrivals will naturally be shifted towards
shorter timescales, thus requiring an accurate determination of the full
first passage statistics. 
Standard approaches focusing on the mean first passage time alone,
would therefore fail in the few encounter limit, whereas 
our new framework provides an accurate and consistent result (see also \cite{hart18_arxiv} and Fig.~2 therein for more details).

\begin{figure}
 \includegraphics{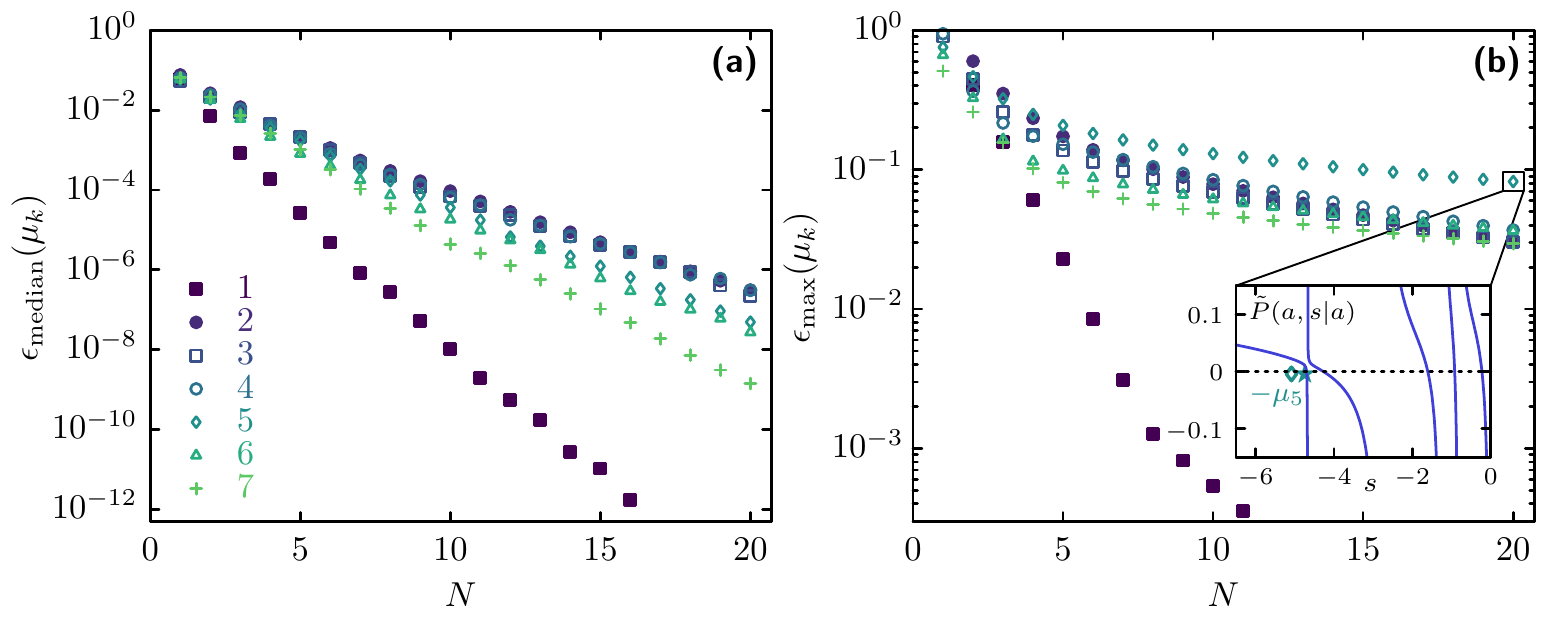}
 \caption{Rate of convergence of the truncated duality solution
   towards the respective numerical solution for randomly generated folding landscapes.
 We generated 100 folding landscapes according to Fig.~\ref{fig:discrete_folding}
 and determined for each model the error of the truncated Newton
 series \eqref{eq:mu_k_newton_N}  in terms of $\epsilon(\mu_k)\equiv |\mu_{k,N}-\mu_k|/\mu_k$ for all first passage eigenvalues $k=1,2,\ldots,7$.
 (a) The typical error given by the median over all respective errors
 $\epsilon(\mu_k)$,  as function of the number of terms $N$ in the
 truncated Newton series \eqref{eq:mu_k_newton_N}. That is,
 50  models generated a smaller error $\epsilon(\mu_k)\le\epsilon_\text{median}(\mu_k)$
 and 50 models
 generated a larger error $\epsilon(\mu_k)\ge\epsilon_\text{median}(\mu_k)$. (b)~Maximal error $\epsilon(\mu_k)\equiv |\mu_{k,N}-\mu_k|/\mu_k$
 out of all 100 randomly generated models as function of $N$.
 Here the maximal error (\emph{worst case}) was observed for the fifth mode.
 The inset shows the diagonal of the propagator $\tilde P(a,s|a)$
 as function of $s$ for the model corresponding to the worst case; the 
 approximations $-\mu_{k,N}$ for the corresponding first passage eigenvalue at $s=-\mu_5$  are indicated for $N=20$ (diamond) and $N=200$ (star).
 }
\label{fig:folding_result2}
\end{figure}

In Fig.~\ref{fig:folding_result2} we systematically analyze the deviation of the truncated
Newton series \eqref{eq:mu_k_newton_N} 
with respect to corresponding numerically obtained first passage eigenvalues $\mu_k$
for 100 randomly generated
folding landscapes. For a given landscape the relative error is quantified in
terms of the dimensionless quantity $\epsilon\equiv|\mu_k^N-\mu_k|/\mu_k$, and Fig.~\mysubref{fig:folding_result2}{a} depicts the typical error
characterized by the median of the individual errors for all seven modes, respectively.
Note that for $N=20$ the relative error of the finite Newton series is typically below $10^{-6}$.

Fig.~\mysubref{fig:folding_result2}{b} displays the maximal error out
of 100 randomly picked landscapes. We observe that larger errors can
occur if a first passage eigenvalue is located immediately after a gap in the relaxation spectrum.
The smaller error of the slowest first passage mode  $\mu_1$
is due to the fact that $\mu_1$ cannot be located after such a gap due to
the interlacing theorem \eqref{eq:interlace}, which implies $\mu_1\le\lambda_1$.
In this specific example the maximum relative error out of 
100 models randomly generated models
is found for the fifth mode ($\mu_5$); the inset of Fig.~\mysubref{fig:folding_result2}{b} shows the corresponding $\tilde P(a,s|a)$ as well as the result form the finite Newton
series with $N=20$ (see diamonds in the inset of Fig.~\mysubref{fig:folding_result2}{b}).
In this extreme scenario the weight of the fifth relaxation mode $\Psi_5\ll \Psi_{l\neq 5} $
is almost negligible compared to other weights, leading to an almost
vanishing weight $w_5$, which would in turn require an increased number
of terms $N$ entering the Newton series. Increasing the number of terms in the truncated Newton series from $N=20$ to $N=200$
reduces the deviation from  $\epsilon\simeq10^{-1}$
to $\epsilon\simeq10^{-2}$, the result for $N=200$ is marked by
the star in the inset of Fig.~\mysubref{fig:folding_result2}{b}.
Fig.~\ref{fig:folding_result2} readily demonstrates that the our
duality can be robustly and reliably applied to
all Markov state models.

\subsection{Ornstein-Uhlenbeck process}
\label{sec:OU}
Let us now consider a linear Ornstein-Uhlenbeck, which corresponds to
a diffusion process in a harmonic potential $\beta U(x)=\omega
x^2/2$. The corresponding Fokker-Planck operator reads $\mL=D\del_x\omega x+D\del_x^2$.
The eigendecomposition of the relaxation process in the absence of
reflecting boundaries is well known. The respective eigenvalues
are given by $\lambda_k=D\omega k$ with the corresponding eigenfunctions  \cite{gard04}
\begin{equation}
 \begin{aligned}
  \psiR_k(x)& \equiv \bra{x}\kket{\psiR_k}=
  \frac{\e^{-\omega x^2/2}}{\sqrt{2\pi/\omega}}
  \frac{H_k(x\sqrt{\omega/2})}{k!2^k},\\
  \psiL_k(x)&\equiv\bra{\psiL_k}\kket{x} =H_k(x\sqrt{\omega/2}),
  \label{eq:Psi_ax}
\end{aligned}
\end{equation}
where $H_k$ is the $k$th Hermite polynomial. Although this process is extremely well studied, a closed-form analytical result for the first passage time distribution $\wp_a(t|x_0)$ for any non-centered target position $a\neq 0$
remained elusive \cite{alil05,greb15,nybe16}.
We note that the well known analytical solution of the Laplace transform of the probability density $\tilde{\wp}_a(s|x_0)$ in terms of Hermite polynomials \cite{alil05,darl53,sieg51,verg18} until now could only be analytically inverted to $\wp_a(t|x_0)$ for the special case $a=0$ \cite{alil05}.
Furtheremore, the exact large deviation limit $\wp_a(t|x_0)\simeq w_1(x_0)\e^{-\mu_1 t}$ was just recently derived in \cite{gode16}.
To obtain the full first passage time distribution we here use Eqs.~\eqref{eq:Fk_def}-\eqref{eq:k*} as follows.
Inserting Eq.~\eqref{eq:Psi_ax} into Eqs.~\eqref{eq:Fk_def}-\eqref{eq:k*}
yields the modifed diagonal of the propagator
\begin{align}
 f(s,k)
& =(s+D\omega k^*)\sum_{l=0}^M
 \frac{\e^{-\omega a^2/2}}{\sqrt{2\pi/\omega}}
  \frac{H_l(a\sqrt{\omega/2})^2}{l!2^l(s+D\omega l)}
\end{align}
where $k^*(k)=k,k-1$ is chosen according to Eq.~\eqref{eq:k*}, which is equivalent to
$f(-\bar{\mu}_{k},k)<0$ with  $\bar{\mu}_k=D\omega (k-1/2)$.
Note that we truncated the sum after $M$ terms for the numerical evaluation, whereas
the exact formal result corresponds to $M=\infty$.
The first line of Eq.~\eqref{eq:fn} is then simply given by $f_0(k)=f(-\bar{\mu}_k,k)$ and the second line of Eq.~\eqref{eq:fn} becomes
\begin{equation}
\begin{aligned}
 f_{n\ge 1}(k)=\frac{\e^{-\omega a^2/2}}{(D\omega)^{n}\sqrt{2\pi/\omega}}\sum_{\substack{l=0\\l\neq k^*}}^M
  \frac{H_l(a\sqrt{\omega/2})^2(l-k^*)}{l!2^l(k-l-1/2)^{n+1}}. 
\end{aligned}
\end{equation}
The $k$th first passage eigenvalue $\mu_k$ is determined by using the finite Newton series \eqref{eq:mu_k_newton_N}, where the almost triangular matrix is taken from Eq.~\eqref{eq_def_An}, and the corresponding first passage weights $w_k(x_0)$
are determined using the residue theorem \eqref{eq:wk_explicit}. Note that our theory allows for the first time to determine analytically all first passage eigenvalues $\{\mu_k\}$ as well as the weights $\{w_k\}$ and, therefore, also provides a complete solution to the first passage time density $\wp_a(t|x_0)$.

\begin{figure}
 \centering
\includegraphics{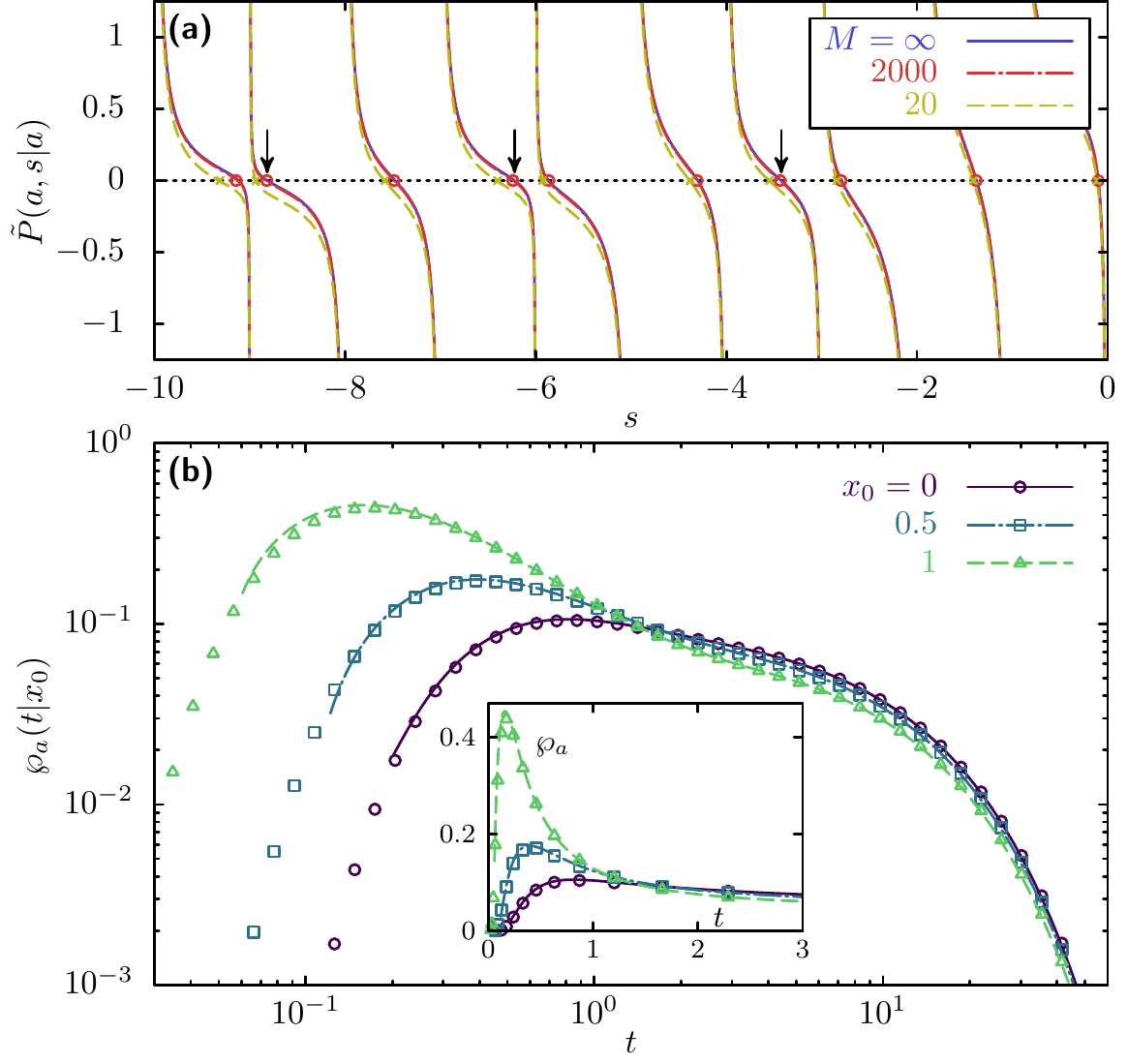}
 \caption{Analytical first passage time density for the
   Ornstein-Uhlenbeck process. (a)~Exact diagonal of the relaxation
   propagator (solid blue line) versus a finite mode expansion  $\tilde{P}(a,s|a)\approx\sum_k^M\Psi_k(a,a)/(s+D\omega k)$ for $M=2000$ (dash-dotted red line) and $m=20$ (dashed yellow line);
 the exact solution is obtained from \eqref{eq:propagator_solution}
 with particular solutions  $v_\pm(a,s)=\e^{-\omega a^2/2}H_{-s/(D\omega)}(\pm a\sqrt{\omega/2})$, where $H_{-s}(x)$
 is the generalized Hermite polynomial.
 The symbols represent the roots $s=-\mu_k$ that are determined from the Newton series
 \eqref{eq:mu_k_newton_N} with $N=10$ using $M=2000$ (open red circles) and $M=20$ (yellow crosses  modes, respectively. The three vertical arrows $s=-\mu_k$
 indicate the that correspond to an absorption from  the right, where $w_k(x_0)=0$ for all $x_0\le a$.
 (b) First passage time distribution for three different initial conditions $x_0=0,0.5,1$. We have used $M=2000$ relaxation modes and $N=30$ and used our analytical forward duality to calculate the lines. The symbols represent a numerical inversion of the Laplace transform of $\tilde{\wp}_a(s|x_0)=\e^{\omega (a^2- x_0^2)/2}H_{-s/(D\omega)}(- x_0\sqrt{\omega/2})/H_{-s/(D\omega)}(- a\sqrt{\omega/2})$ according to Ref.~\cite{alil05}.
 Parameters: $a=2$, $\omega=D=1$.
  }
  \label{fig:harmonic}
\end{figure}

Fig.~\ref{fig:harmonic} depicts the results for the case, where the
absorbing point is set at $a=2$. Note that this scenario does not yet
correspond to the well-known high barrier Kramers regime. 
In Fig.~\mysubref{fig:harmonic}{a} we compare the exact $\tilde{P}(a,s|a)$ (solid blue line) with the finite approximation from Eq.~\eqref{eq:Pxx0_FP_M} using $M=20$ (dashed yellow line) and $M=2000$ (dash-dotted red line) relaxation modes, respectively.
The symbols represent the corresponding first passage modes ($s=-\mu_1,-\mu_2,\ldots$).
Using only a small number of relaxation modes $M=20$ (see yellow
crosses) the zeros differ substantially from the respective
numerically obtained solution, which becomes, however, rather well approximated if we increase
the number of modes to $M=2000$ (see open red circles).
We note that such deviations of the first passage modes become particularly
inconvenient
for the modes that are marked by the vertical arrows in Fig.~\mysubref{fig:harmonic}{a}. These first passage modes correspond to an absorption from the right, where
the corresponding weights  vanish $w_k(x_0)=0$ (here $k=4,7,9,\ldots$) completely for all $x_0\le a$, which, however, is only obtained in the limit $M\to\infty$.

This numerical truncation problem can be avoided completely if the relaxation process is considered with a 
reflecting boundary condition as explained in Sec.~\ref{sec:reflecting_wall},
which automatically removes beforehand all zeros marked by the arrows in Fig.~\mysubref{fig:harmonic}{a} ($s=-\mu_4,-\mu_7,-\mu_9,\ldots$).
Nevertheless, to illustrate the power and robustness of our duality approach we
proceed here without a reflecting wall and use $M=2000$.
In  Fig.~\mysubref{fig:harmonic}{b} we show the first passage time distribution
on a log-log scale (see inset for a linear scale) for three different starting positions $x_0=0,0.5,1$ (absorbing point $a=2$). The lines represent the first passage time distribution which is determined using our new method (with $M=2000$
relaxation modes) and the symbols represent the results $\wp_a$ of a
numerical Laplace inversion of the renewal theorem (see figure caption for more details).
We find a perfect agreement between our new analytical method (lines)
and the numerical solution.
For comparison, we imposed a reflecting wall at the target in a related article \cite{hart18_arxiv} and obtained a similarly excellent agreement
between the duality solution and the simulated first passage time
density using a total of $M=40$ relaxation modes to quantify the first
passage time statistics for diffusion in a multi-well potential.
In either case, our new duality framework is exact for infinite $M$,
and hence the desired precision can be tuned at will.

\section{Concluding perspectives}
\label{sec:conclusion}
We rigorously established a duality between the relaxation and the
corresponding first passage processes
in terms of an interlacing of eigenvalues.  In other words, the time-scales at which a particle is
absorbed into the target are proven to interlace with the corresponding relaxation timescales. This duality allows us to understand first passage processes, both qualitatively and quantitatively, in terms of relaxation eigenmodes. For example,  spectral gaps in the relaxation spectrum
translate directly into spectral gaps in the first passage spectrum.
More explicitly, in effectively  one dimensional systems $N$ gaps in the relaxation spectrum,
arising from $N$ local (free) energy basins, translate into $N-1$ gaps in the first passage
time spectrum corresponding to the $N-1$ barriers separating the minima.
Most importantly, we established a duality that allows,  for the first time,
to determine exactly the first passage time distribution from
the corresponding relaxation spectrum.

Our theory is developed end tested
on both, continuous reversible Fokker-Planck dynamics and Markov state jump
processes in arbitrary dimensions.
For convenience and without loss of generality, we restricted the
applications of the duality for systems obeying Fokker-Planck dynamics
to effectively one dimensional problems.
An extension to more general models, for example, to diffusion on graphs would be straightforward, albeit rendering the calculations more cumbersome.

We tested and applied our theory to a discrete Markov state model 
of a simple protein folding landscape and the Ornstein-Uhlenbeck process, while
a continuous analogue of a  folding landscape are discussed elsewhere \cite{hart18_arxiv}.
Notably, we have derived, to the best of our knowledge, for the first time an exact and explicit analytical expression for the first passage time distribution of the Ornstein-Uhlenbeck process.

Looking forward it will be interesting and relevant to apply the
duality to the analysis of first passage processes on
graphs. Applications of the duality to narrow escape problems in
arbitrary dimensions \cite{sing06,schu07,rein09,pill10,isaa16,greb17}
will also be carried out in future studies. 

Finally, an extension of the framework to periodically or constantly driven
systems (i.e., irreversible Markovian dynamics), which goes beyond the long time limit that is presented
in \ref{sec:A:mu1},
will be particularly
challenging. Namely, there the interlacing theorem
cannot  be expected to hold anymore, since both eigenvalue spectra $\{\lambda_k\}$ and $\{\mu_k\}$ can become complex valued.

\appendix
\section{Explicit formula for principal eigenvalue}
\label{sec:A:mu1}
In this appendix we simplify Eqs.~\eqref{eq:Fk_def}-\eqref{eq:k*} in the limit of a time-scale separation and for rare-event asymptotics for the principal
first passage eigenvalue $\mu_1$.
We obtain a compact asymptotic expression of the principal first passage eigenvalue
$\tilde\mu_1\simeq\mu_1$, which is particular accurate if the time-scale of the slowest first passage eigenvalue is well separated from the time-scale of the slowest relaxation
mode ($\mu_1\ll\lambda_1$),
which \textit{inter alia}
refines a previously proposed approximate link between the mean first passage time and the slowest relaxation mode \cite{schu79,matk81}.

First, we redefine Eq.~\eqref{eq:fsk} by setting 
$k=1$, $k^*=0$ and $\bar\mu_1=0$
\begin{equation}
 f(s)=P_\text{eq}(a)+\sum_{l\ge 1}\bra{a}\kket{\psiR_l}\bra{\psiL_l}\kket{a}\frac{s}{s+\lambda_l},
\end{equation}
where we dropped for convenience any argument with $k$ since $k=1$
is assumed throughout this appendix.
The $n$th derivative of $f$ at $s=0$
simplifies with Eq.~\eqref{eq:fn} to
\begin{equation}
 f_n=
 \left\{\begin{array}{ll}
 P_\text{eq}(a)&\text{if $n=0$},\\
n!(-1)^{n+1}\sum_{l\ge1} \bra{a}\kket{\psiR_l}\bra{\psiL_l}\kket{a}/\lambda_l^n &\text{if $n\ge1.$}
 \end{array}\right.
\label{eq:A:fn}
\end{equation}
The almost triangular matrices Eq.~\eqref{eq_def_An} become
\begin{equation}
 \boldsymbol{\mathcal{A}}_n^{i,j}=\frac{f_{i-j+2}\Theta(i-j+1)}{(i-j+2)!}
 \Big[n(i-j+1)\Theta(j-2)
 +i\Theta(1-j)+j-1\Big],
 \label{eq:A:An}
\end{equation}
where we have replaced $f_n(k)$ by $f_n$ from Eq.~\eqref{eq:A:fn}.
Consequently, the Newton series \eqref{eq:mu_k_newton}
also simplifies to
\begin{equation}
 \mu_1=\sum_{n=1}^\infty\frac{f_0^n}{f_1^{2n-1}}\frac{\det\boldsymbol{\mathcal{A}}_n}{(n-1)!}
 \label{eq:A:Newton}.
\end{equation}

\begin{figure}
 \centering
 \includegraphics{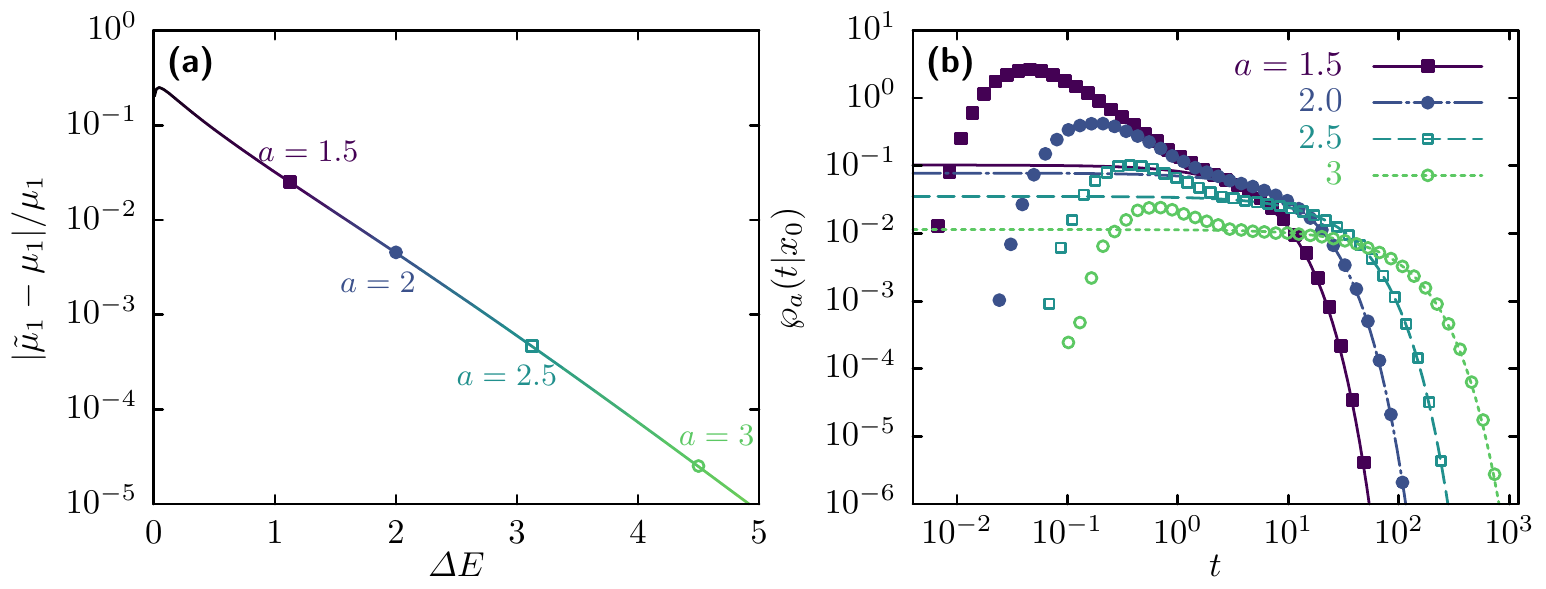}
  \caption{Principal eigenvalue for Ornstein-Uhlenbeck with potential $\beta U(x)=x^2/2$ (with $\omega=D=\beta=1$).
  (a)~Deviation of the approximation $\tilde\mu_1$ from the exact
    first passage eigenvalue $\mu_1$ as function of the height of the energy barrier $\varDelta E=a^2/2$.
  The symbols corresponds to the absorbing points $a=1.5, 2,2.5,3$ used in (b).
  The colored solid line is calculated with Eq.~\eqref{eq:mu1_a}.
  (b)~First passage time density $\wp_a(t|x_0)$ for particle starting from $x_0=1$ as function of time $t$ for target positions
  $a=1.5, 2,2.5,3$. The lines correspond to long time limit approximation
  $\tilde  w_1(x_0)\tilde{\mu}_1\exp(\tilde{\mu}_1 )\simeq \wp_a(t|x_0)$.
  The symbols represent $\wp_a(t|x_0)$ deduced from a histogram over $10^6$ simulated trajectories. The weight $w_1(x_0)$ is deduced from the first line of
  Eq.~\eqref{eq:wk_explicit}, where $\mu_1$ is replaced by $\tilde{\mu}_1$ and we have inserted the propagator from
  \eqref{eq:propagator_solution} with the solutions $v_\pm(x,s)=\e^{-x^2/2}H_s(\pm x/\sqrt{2})$.}
  \label{fig:mua}
\end{figure}

If we now set $f_{3}=f_{4}=\ldots=0$  in the almost triangular matrices \eqref{eq:A:An}, that is $\tilde{\boldsymbol{\mathcal{A}}_n}\equiv\boldsymbol{\mathcal{A}}_n|_{f_{3}=f_{4}=\ldots=0}$,
the resulting matrix $\tilde{\boldsymbol{\mathcal{A}}_n}$ becomes triangular,
implying that its determinant is simply given by the
product of the diagonal elements
\begin{align}
 \det\tilde{\boldsymbol{\mathcal{A}}_n}&=\prod_{i=1}^{n-1}\boldsymbol{\mathcal{A}}_n^{i,i}
=(f_2/2)^{n-1}\frac{(2n-2)!}{n!},
\end{align}
where we have inserted Eq.~\eqref{eq:A:An} and evaluated the product
in the last step.
Replacing $\boldsymbol{\mathcal{A}}\to\tilde{\boldsymbol{\mathcal{A}}_n}$
in the
Newton series \eqref{eq:A:Newton} finally
yields exactly
\begin{equation}
 \tilde{\mu}_1\equiv \sum_{n=1}^\infty\frac{f_0^n}{f_1^{2n-1}}\frac{ \det\tilde{\boldsymbol{\mathcal{A}}_n}}{(n-1)!}=\frac{f_1-\sqrt{f_1^2-2f_0f_2}}{f_2}.
 \label{eq:mu1_a}
\end{equation}
Eq.~\eqref{eq:mu1_a} is nothing but the
root of the second order Taylor expansion of $f(s)$ around $s=0$ (i.e., the parabolic equation).
This approximation is quite accurate whenever $ \tilde{\mu}_1\ll \lambda_1$.

If the target is located at a high energy barrier,
such that slowest first passage eigenvalue is exponentially suppressed by the 
(free) energy at the target (i.e.,  $\mu_1\propto\e^{-U(a)}$), Eq.~\eqref{eq:mu1_a}
will lead to a quite accurate approximation $\tilde{\mu}_1$, which can be seen in Fig.~\ref{fig:mua}.
More precisely, in Fig.~\mysubref{fig:mua}{a} we depict the relative error $|\tilde\mu_1-\mu_1|/\mu_1$
as function of the target-site energy $\varDelta E =a^2/2$
for the Ornstein-Uhlenbeck process from Sec.~\ref{sec:OU}
with $U(x)=x^2/2$ ($\omega=D=\beta=1$).
Conversely, Fig.~\mysubref{fig:mua}{b} displays the results the first passage time distribution
to four different target positions
$a=1.5,2,2.5,3$ for a particle starting from $x_0=1$.
The symbols represent histograms for $\wp_a(t|x_0)$ deduced from $10^6$ Brownian dynamics trajectories,
and the lines correspond to the large deviation asymptotic
$\wp_a(t|x_0)\simeq w_1(x_0)\tilde\mu_1\e^{-\tilde \mu_1 t}$ deduced from \eqref{eq:mu1_a}.
We conclude that the limit  $\mu_1\ll\lambda_1$ lead to both, a quite accurate approximation $\tilde\mu_1\simeq\mu_1$
and to an effectively single exponential decay $\mu_1\ll\mu_2$ of the first passage statistics, which extends previous results \cite{bere04} (see also \cite{bico00}).

Moreover, a spectral gap such as $\lambda_1\ll\lambda_2$ 
will also render $f_{n\ge 2}$ from Eq.~\eqref{eq:A:fn} to be negligibly small
if the target is not located at the global minimum of the potential. For example,
a multi-barrier crossing, as the one studied in Ref.~\cite{hart18_arxiv} (see Fig.~5 therein),
the principal first passage eigenvalue from Eq.~\eqref{eq:mu1_a}
deviates less than two percent from the exact value $\mu_1$, i.e.,
$|\mu_1-\tilde\mu_1|/\mu_1<0.02$.
Notably, the approximation
\eqref{eq:mu1_a}
refines previous conjectures that the mean first passage time to escape from
the deepest potential basin corresponds to the first nonzero relaxation mode \cite{schu79,matk81}.

We note that Eq.~\eqref{eq:mu1_a} can be reformulated to give
\begin{equation}
 \tilde\mu_1=\frac{\sigma_1}{2\sigma_2}\left[\sqrt{1+4\frac{P_\text{eq}(a)\sigma_2}{\sigma_1^2}}-1\right],
\end{equation}
where we inserted $f_0=P_\text{eq}(a)$ and defined $\sigma_n=\sum_{l\ge1} \bra{a}\kket{\psiR_l}\bra{\psiL_l}\kket{a}/\lambda_l^n$. This relation is equivalent to Eq.~(17)
from a related article \cite{hart18_arxiv}.
 
Finally, we emphasize that Eqs.~\eqref{eq:A:Newton} and \eqref{eq:mu1_a} apply also to  homogeneous irreversible Markov processes, i.e.,  the relations from this appendix are not restricted
to hold just for reversible Markov chains.

\ack

We thank Matteo Polettini for useful comments on our manuscript.
The financial support from the German Research
Foundation (DFG) through the Emmy Noether Program
``GO 2762/1-1'' (to AG) is gratefully acknowledged.

\section*{References}

\providecommand{\href}[2]{#2}\begingroup\raggedright\endgroup

 \end{document}